\documentclass[letterpaper,twocolumn,10pt]{article}
\usepackage{usenix-2020-09}

\usepackage{tikz}
\usepackage{amsmath}
\usepackage[absolute]{textpos}

\def\noeditingmarks{}
\clearpage{}%
\usepackage{lipsum}
\usepackage{booktabs}  %
\usepackage{dcolumn}   %
\usepackage{multirow}  %
\usepackage{makecell}  %
\usepackage{float}     %
\usepackage{bbm}       %
\usepackage{rotating}
\usepackage{xspace}
\usepackage{hyphenat} 
\usepackage{enumerate}
\usepackage{mfirstuc} %
\usepackage{tikz} %
\usepackage{xparse} %
\usepackage{listings}
\definecolor{dkgreen}{rgb}{0,.6,0}
\definecolor{dkblue}{rgb}{0,0,.6}
\definecolor{dkyellow}{cmyk}{0,0,.8,.3}
\usepackage{enumitem}
\setlist{leftmargin=0.2in}

\usepackage{wrapfig}
\usepackage{textcomp}
\usepackage{tabularx}
\usepackage{pifont}
\usepackage{url}

\usepackage{breakurl}

\usepackage{wrapfig}

\usepackage{ulem}
\usepackage{mfirstuc}

\usepackage{colortbl} %
\usepackage{array}    %

\usepackage{dblfloatfix}

\usepackage[font=small]{caption}
\usepackage{subcaption}

\usepackage[noend]{algpseudocode}
\algrenewcomment[1]{\hfill// #1}%
\algrenewcommand\algorithmicindent{1.25em}

\definecolor{darkgreen}{rgb}{0,0.75,0}

\algrenewcommand\algorithmicdo{:}
\algrenewcommand\algorithmicwhile{\textbf{while}}
\algrenewcommand\algorithmicfor{\textbf{for}}
\algrenewcommand\algorithmicforall{\textbf{for all}}
\algrenewcommand\algorithmicloop{\textbf{loop}}
\algrenewcommand\algorithmicrepeat{\textbf{repeat}}
\algrenewcommand\algorithmicuntil{\textbf{until}}
\algrenewcommand\algorithmicprocedure{\textbf{procedure}}
\algrenewcommand\algorithmicfunction{\textbf{function}}
\algrenewcommand\algorithmicif{\textbf{if}}
\algrenewcommand\algorithmicthen{:}
\algrenewcommand\algorithmicelse{\textbf{else}}
\algrenewcommand\algorithmicrequire{\textbf{Require}:}
\algrenewcommand\algorithmicensure{\textbf{Ensure}:}
\algrenewcommand\algorithmicreturn{\textbf{return}}
\algdef{SE}[WHILE]{While}{EndWhile}[1]{\algorithmicwhile\ #1\algorithmicdo}{\algorithmicend\ \algorithmicwhile}%
\algdef{SE}[FOR]{For}{EndFor}[1]{\algorithmicfor\ #1\algorithmicdo}{\algorithmicend\ \algorithmicfor}%
\algdef{S}[FOR]{ForAll}[1]{\algorithmicforall\ #1\algorithmicdo}%
\algdef{SE}[IF]{If}{EndIf}[1]{\algorithmicif\ #1\algorithmicthen}{\algorithmicend\ \algorithmicif}%
\algdef{C}[IF]{IF}{ElsIf}[1]{\algorithmicelse\ \algorithmicif\ #1\algorithmicthen}%
\algnotext{Function}                
\algnotext{EndFunction}
\algnotext{EndFor}
\algnotext{EndIf}
\algnotext{EndWhile}
\algnewcommand\Or{\textbf{or}\xspace}
\algnewcommand\myAnd{\textbf{and}\xspace}

\usepackage{amsmath,amscd}
\usepackage{amssymb}
\usepackage{amsfonts}
\usepackage{amsthm}

\usepackage[noeka]{mathrmletter}
\usepackage{tgtermes}

\newif\ifextended
\newif\iflongbatching
\newif\ifsubmission
\newif\ifelementary
\newif\ifwithdbproof
\newif\ifbuildanonapdx

\newenvironment{myitemize2}%
  {\begin{list}{\labelitemi}{\itemsep1pt \topsep2pt \parsep0.00in
  \partopsep=0pt \leftmargin1.2em}}%
  {\end{list}}

\let\latexusecounter=\usecounter
\def\compactsortof{\itemsep=0in \topsep=2pt \parsep=0.00in \partopsep=0pt
\leftmargin=1.2em}
\newenvironment{myenumerate2}
  {\def\usecounter{\compactsortof\latexusecounter}
   \begin{enumerate}}
  {\end{enumerate}\let\usecounter=\latexusecounter}

\ifx\noeditingmarks\undefined
   \newcommand{\pgwrapper}[3]{\begingroup \color{#1} #2: #3 \endgroup}
   \newcommand{\pgwrapperb}[1]{\textbf{#1}}
   \newcommand{\dangerwrapper}[1]{{\color{red}#1}}
\else
   \newcommand{\pgwrapperb}[1]{}
   \newcommand{\pgwrapper}[3]{}
   \newcommand{\dangerwrapper}[1]{}
\fi

  \def\hn{\sffamily\selectfont}
  \newcommand{\mpfont}{\hn\scriptsize}

  \ifx\noeditingmarks\undefined
      \newcommand{\MPworker}[2]{\unskip{\color{#1}\vrule\vrule}{\marginpar{\raggedright\color{#1}\mpfont #2}}}
  \else
    \newcommand{\MPworker}[2]{\unskip}
\fi

\newcommand{\ZC}[1]{\MPworker{blue}{ZC: #1}}

\ifx\noeditingmarks\undefined
    
\else
    
\fi

\def\t{\textit}

\newcommand{\CF}[1]{\xmakefirstuc{#1}}
\newcommand{\heading}[1]{
  \vspace{1ex}
  \noindent
  \textbf{#1}}

\newcommand{\prob}{the Reconfigurable Machine Scheduling Problem\xspace}
\newcommand{\shortprob}{RMS\xspace}

\newcommand{\rules}{reconfiguration rules\xspace}
\newcommand{\oneslice}{1/7 instance\xspace}
\newcommand{\twoslice}{2/7 instance\xspace}
\newcommand{\threeslice}{3/7 instance\xspace}
\newcommand{\fourslice}{4/7 instance\xspace}
\newcommand{\sevenslice}{7/7 instance\xspace}
\newcommand{\optimizer}{optimizer\xspace}
\newcommand{\Optimizer}{\CF{\optimizer}}
\newcommand{\scheduler}{controller\xspace}
\newcommand{\Scheduler}{\CF{\scheduler}}
\newcommand{\deployment}{deployment\xspace}
\newcommand{\deployments}{\deployment{}s\xspace}
\newcommand{\slos}{SLOs\xspace}
\newcommand{\slo}{SLO\xspace}
\newcommand{\fullslos}{service-level objectives\xspace}
\newcommand{\plan}{transition plan\xspace}
\newcommand{\exchangeandcompact}{exchange-and-compact\xspace}
\newcommand{\algotwo}{\exchangeandcompact}
\newcommand{\sstate}{completion rates\xspace}
\newcommand{\gconfig}{GPU configuration\xspace}
\newcommand{\gconfigs}{\gconfig{}s\xspace}
\newcommand{\utility}{utility\xspace}
\newcommand{\utilities}{utilities\xspace}
\newcommand{\serv}[1]{$\textrm{service}_{#1}$}
\newcommand{\gpusevenseven}{A100-7/7\xspace}
\newcommand{\gpuoneseven}{A100-7$\times$1/7\xspace}

\newcommand{\opreconf}{$op_{reconf}$\xspace}
\newcommand{\rulereconf}{$rule_{reconf}$\xspace}

\newcommand{\modelone}{\texttt{densenet121}\xspace}
\newcommand{\modeltwo}{\texttt{xlnet-large-cased}\xspace}

\newcommand{\varcr}{\t{comp\_rates}}
\newcommand{\vartrue}{\t{True}}

\newcommand{\varconf}{\t{conf}}

\clearpage{}%

\include{vars}

\newcommand{\sys}{\textsc{MIG-serving}\xspace}
\newcommand{\Sys}{\CF{\sys}\xspace}

\begin{document}

\date{}

\title{\Large \bf Serving DNN Models with Multi-Instance GPUs: \\A Case of the Reconfigurable Machine Scheduling Problem}

\author{\rm {
    Cheng Tan$^{\dagger\star}$,
    Zhichao Li$^\dagger$,
    Jian Zhang$^{\dagger\star}$,
    Yu Cao$^{\dagger\circ}$,
    Sikai Qi$^\dagger$,
    Zherui Liu$^\dagger$,
    Yibo Zhu$^\dagger$,
    Chuanxiong Guo$^\dagger$}\\
  \emph{$^\dagger$ByteDance Inc.} \quad
  \emph{$^\star$Northeastern University} \quad \emph{$^\circ$New York University}
  }

\maketitle
\frenchspacing

\begin{abstract}
Multi-Instance GPU (MIG) is a new feature introduced by NVIDIA A100 GPUs
that partitions one physical GPU into multiple %
GPU instances.
With MIG, %
A100 can be the most cost-efficient GPU \textit{ever}
for serving Deep Neural Networks (DNNs). %
However, discovering the most efficient GPU partitions is challenging.
The underlying problem is NP-hard;
moreover, it is a new abstract problem,
which we define as \textit{\prob} (RMS).

This paper studies serving DNNs with MIG, a new case of RMS. 
We further propose a solution, \sys.
\sys is an algorithm pipeline that blends a variety of newly designed algorithms
and customized classic algorithms,
including
a heuristic greedy algorithm, %
Genetic Algorithm (GA),
and Monte Carlo Tree Search algorithm (MCTS).
We implement \sys on Kubernetes.
Our experiments show that
compared to using A100 as-is, %
\sys can save up to 40\% GPUs
while providing the same throughput.

\end{abstract}

\section{Introduction}
\label{s:intro}

NVIDIA A100~\cite{a100} is the latest
and the most powerful GPU 
launched in 2020.
Seemingly, A100 is not cost-efficient for DNN serving (inference)
because serving may not fully utilize GPU resources.
However, we argue that, equipped with
a new feature---\textit{Multi-Instance GPU}---A100 can be
the most cost-efficient GPU \textit{ever} for DNN serving.
\ZC{NVIDIA is releasing A30 with MIG, specifically for inference. It may be more
cost effective than A100. We probably want to cover this info. }

Multi-Instance GPU (MIG) is a new hardware feature
introduced by A100.
MIG allows people to partition one physical GPU
into some number of \textit{GPU instances} (or \textit{instance} for short)
that are %
hardware isolated.
For example, an A100 can be partitioned up to 7 instances,
and each instance has its own 
processors,
memory, L2 cache, and bus bandwidth.
Moreover, small instances can be merged into larger instances,
for example, two of the 7 instances in A100 (which we call \textit{\oneslice{}s})
can merge to a \textit{\twoslice} with twice the resources.

\begin{figure}[t]
\begin{center}
\includegraphics[width=0.48\textwidth]{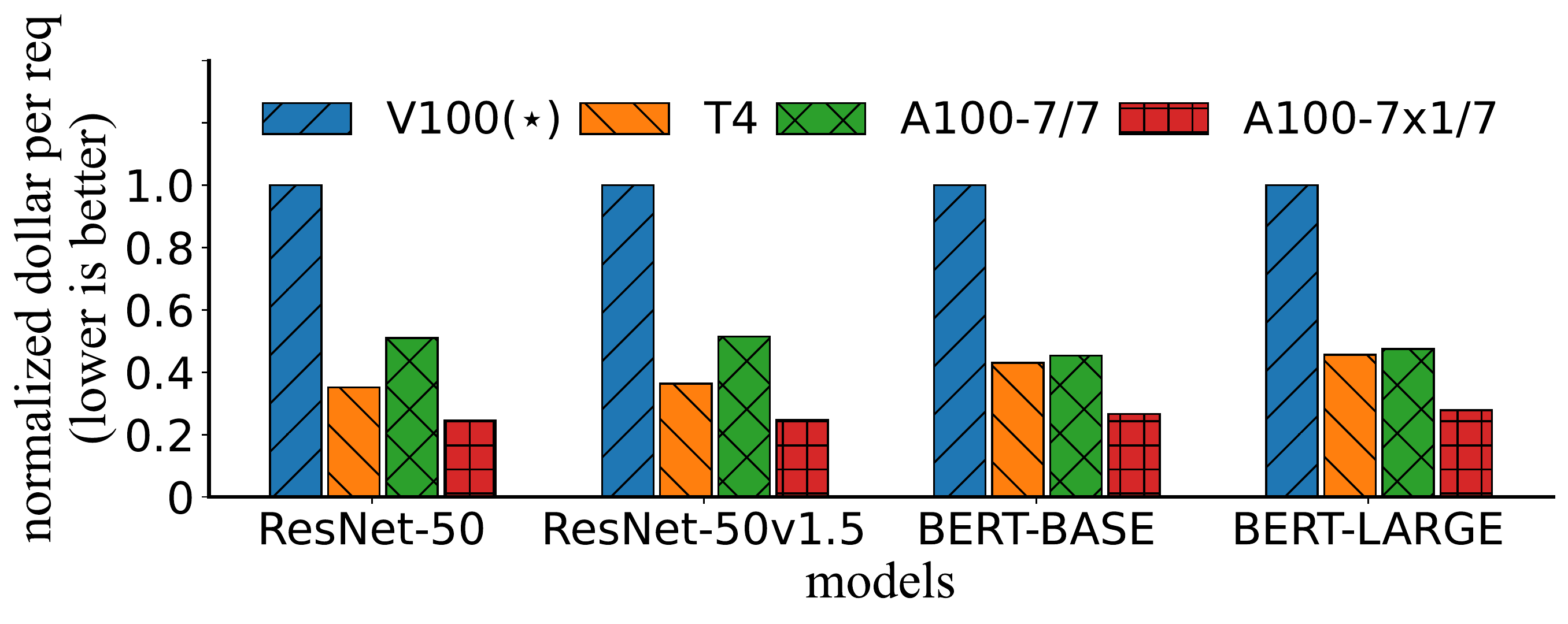}
\end{center}
  \vspace{-3ex}
\caption{Normalized cost per request for different DNN
models (batch size 8, INT8, TensorRT 7.2)
on different GPUs.
The cost is calculated based on
model serving performance from NVIDIA inference benchmarks~\cite{nvidia_perf}
and the price from AWS~\cite{aws_v100, aws_t4, aws_a100}.\\
$(\star)$: NVIDIA does not provide inference performance of INT8
for the three leftmost models on V100~\cite{nvidia_perf};
they provide (and we use) ``Mixed'' precision instead.}
\label{fig:motivation}
  \vspace{-3ex}
\end{figure}

To understand the serving costs (in dollars) on different GPUs,
we calculate how much one needs to pay for serving one request
using varied GPUs on AWS~\cite{aws_v100, aws_t4, aws_a100},
including V100, T4, and A100---in which A100 is configured into two variants:
using A100 as a
whole (\gpusevenseven) and partitioning A100 into seven \oneslice{}s
(\gpuoneseven).
\footnote{Note that
we haven't considered A30 (another MIG-enabled GPU) yet
because no cloud deploys A30~\cite{nvidia_cloud}
hence we are unable to compare price fairly.}
Figure~\ref{fig:motivation} shows the result:
\gpuoneseven is the most
cost-efficient setup for all models.

Can we do better than \gpuoneseven?
The answer is yes.
We observe that different models
have different preferences about instance sizes (\S\ref{s:study}),
thus we can improve inference performance by leveraging A100's heterogeneity;
namely, partitioning an A100 into different sized instances,
for example, a \fourslice, a \twoslice, and a \oneslice.

Meanwhile, however,
heterogeneity improve efficiency at the cost of simplicity.
It raises many questions (also opportunities), just to name a few:
how to partition GPUs regarding instances of different sizes?
Should we mix different models in a GPU, and which ones to mix?
DNN service deployers have different throughput and latency requirements for different models
(defined as \textit{service level objectives}, \textit{\slos}).
Consequently, the GPU configuration that has the highest throughput per resource
is not necessarily the best choice.
How can we reflect \slos in the GPU configurations? 

All these questions lead to our core question,
\textit{how to configure MIG-enabled GPUs to most efficiently meet \slos?}
By ``most efficiently'', we mean that
GPUs serving DNN models (called \textit{services}) %
can satisfy \slos
with the minimum number of GPUs.
Our problem has three characteristics
which in combination make the problem novel and challenging.

First, different DNNs have different performance per resource
on different sized instances (\S\ref{s:study}).
This means that 
we cannot simply assume that two \oneslice{}s equal one \twoslice
and assign resources by total amounts,
which is a common assumption used by traditional resource allocation algorithms (like allocating CPU cores).

Second, instance allocation is restricted:
partitioning GPUs follows specific and (arguably) peculiar rules.
These rules may reject seemingly valid partitions.
For example,
an A100 cannot allocate a \threeslice when having a running \fourslice,
even if it has three free units of resources.
This ``no $4/7 + 3/7$'' is a \textit{hard-coded rule} (\S\ref{s:migintro}),
\ZC{no 3/7 out of 4/7, and no 3/7 + 4/7, are two different things. Maybe more clearer. } which has something to do with the hardware overhead of MIG~\cite{a100_doc}.
These rules break an assumption made by many resource allocators (like memory and disk allocators)
that having $n$ units of free resources indicates one can always allocate a chunk of $n$
resources (by some rearrangements, if needed).

Third, MIG supports \textit{partial reconfiguration}~\cite{steiger04operating}:
a subset of a GPU's instances can be repartitioned on-the-fly,
without affecting other working instances on the same GPU.
Partial reconfiguration differs from classic reconfigurable setup
(like RMTs~\cite{landers01reconfigurable})
because the amount of resources involved in
one reconfiguration is a \textit{variable},
whereas classic
reconfigurable devices, like RMTs, have a basic reconfigurable unit
which is \textit{fixed} in size.

\bigskip

We define an abstract problem, \textit{\prob},
that captures and formally specifies
the above three characteristics.
The problem is NP-hard (\S\ref{s:realproblem}).
Despite being computationally expensive to solve,
the problem is crucial for deep learning tasks running on MIG-enabled GPUs,
as the potential of MIG is enormous.
In our experiments,
we can save up to 40\% GPUs
by carefully configuring MIG
instead of ignoring MIG and using GPUs as a whole (\S\ref{s:eval}).

This paper describes a system called \sys, which
aims at serving DNNs with MIG.
\Sys takes DNN models and their \slos as inputs, %
and produces a set of GPU partitions and service assignments,
called a \textit{\deployment},
that satisfies all \slos
and uses as few GPUs as possible.

\Sys consists of two main components: \textit{\optimizer} and \textit{\scheduler}.
\Optimizer is responsible for generating and optimizing \deployments.
Specifically, it can generate a valid \deployment quickly (in minutes);
while if more time and computing resources are available,
\optimizer can gradually improve the result.
\Scheduler is in charge of actually applying the \deployment to GPU clusters.
In the this process,
\scheduler ensures that end users will not experience service interruptions.

The contributions of this paper are as follows:

\begin{myitemize2}

\item \textit{A study of model serving performance with MIG (\S\ref{s:study}, Appendix~\ref{appx:study}).}
We study 49 trained models from PyTorch Hub~\cite{pytroch_hub} and TensorFlow Hub~\cite{tensorflow_hub}, and evaluate their performances on different sized instances.
We observe that the throughput of most models does not grow linearly with the increase of resources.

\item \textit{Definition of \prob (\S\ref{s:prob}).}
We define the problem in theoretical terms
to highlight the fundamental difficulties and
the relationship to other classic scheduling problems.

\item \textit{An algorithm pipeline for serving DNNs with MIG
(\S\ref{s:algoone}, \S\ref{s:algotwo})}. We design a two-step pipeline that
explores \gconfigs and searches for cost-efficient \deployments, including:

  \begin{myenumerate2}

  \item \textit{\Optimizer: balancing two conflicting requirements (\S\ref{s:algoone}).}
  \Optimizer needs to search for a \deployment that satisfies \slos.
  Except being computationally expensive,
  this search has two conflicting requirements in practice:
  (a) discovering a valid \deployment quickly and
  (b) pursuing the most efficient \deployment.
  We tackle this challenge by a two-phase algorithm that combines two
  ``template algorithms''---a fast algorithm and a slow algorithm---through a
  tailored Genetic Algorithm (GA).

 \item \textit{\Scheduler: transparent \deployment transition (\S\ref{s:algotwo}).}
 From time to time, services get updated
 and \optimizer produces new \deployments to reflect the changes.
 \Scheduler is required to execute the \deployment transitions
 \textit{transparently}---without affecting user experiences.
 To achieve such transparency,
 \scheduler uses an algorithm, \textit{\algotwo}, which
 guarantees that during transitions, service throughputs
 are always greater than the required throughputs of the new or old \deployments,
 whichever is smaller.

  \end{myenumerate2}

\item \textit{A built system and experimental evaluation (\S\ref{s:impl}, \S\ref{s:eval})}.
We implement \sys on Kubernetes and experiment with it on a 24 A100 GPU cluster.
\Sys can save up to 40\% of GPUs compared to using A100 disabling MIG (\S\ref{s:eval:algos}).
Also, \sys is able to finish \deployment transitions
between two real-world workloads within half an hour (\S\ref{s:eval:transition}).

\end{myitemize2}

\section{Multi-Instance GPU}
\label{s:mig}

This section introduces
MIG in detail (\S\ref{s:migintro})
and studies the performance characteristics of DNN models running
on different sized instances (\S\ref{s:study}).
We further describe two straightforward approaches to use MIG for DNN inferences
(\S\ref{s:homoserving}),
which will serve as baselines in our experiments (\S\ref{s:eval}).

\subsection{NVIDIA A100 MIG}
\label{s:migintro}

MIG is a hardware feature %
that allows users to partition a GPU into
multiple \textit{GPU instances} (or \textit{instance} for short). Each instance functions as a traditional GPU.
Current A100 GPU implementation has 7 slices of resources\footnote{The
    description of ``7 slices'' is a simplification;
    in fact, different resource categories are assigned differently.
    For example, a \oneslice has 1/7 processor resources, but only 1/8 memory of A100.
    See A100 whitepaper~\cite{a100_doc} for more information.}
and people can organize these resources
in many ways with diverse sized instances.
For example, a GPU can be partitioned into three instances
with 1/7, 2/7, and 4/7 of the total resources respectively.
In the rest of the paper, we call an instance with 1/7 of total resources
as a \oneslice (similarly for instances of other sizes).

Different from resource sharing like MPS (Multi-Process Service),
MIG's instances do not share computational resources:
instances have separate streaming multiprocessors (SM), GPU memory, and L1/L2 cache.
In addition, instances provide fault and performance isolation
by having dedicated on-chip crossbar ports, L2 cache banks,
memory controllers, and DRAM address buses.
Essentially, an instance is a full-fledged GPU,
except some of them are packed in the same
``metal box'' (an A100).

As mentioned earlier (\S\ref{s:intro}),
MIG's instance allocation follows specific rules;
hence having $n$ units of free resources does not
imply %
that a GPU is able to allocate an $n$/7 instance.
On the one hand, resources can only be grouped into specific sized
instances---1/7, 2/7, 3/7, 4/7, 7/7 instances,
whereas others (5/7 and 6/7 instances) are not allowed.
On the other hand, %
the occupied resources also influence the
possible allocations.
As an example,
for a GPU with two running \threeslice{}s,
allocating a \oneslice is prohibited.

\begin{figure}[t]
\begin{center}
\includegraphics[width=0.48\textwidth]{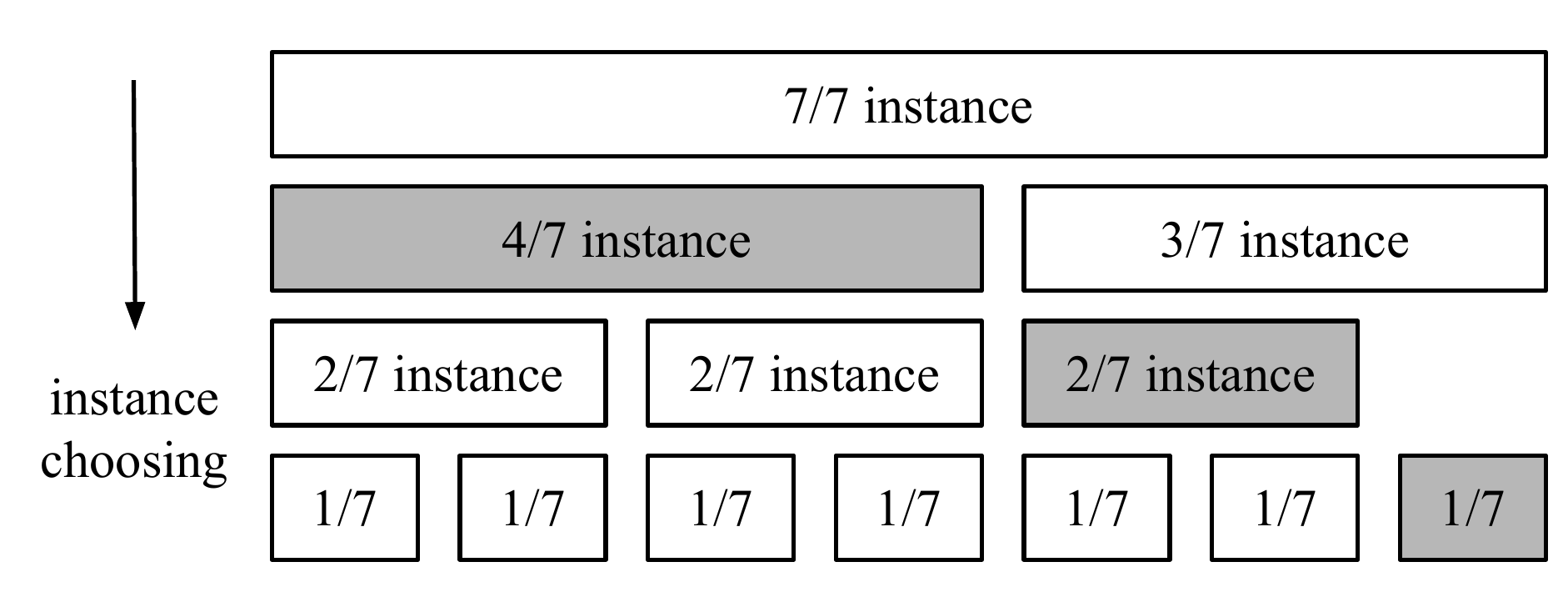}
\end{center}
  \vspace{-3ex}
\caption{Instance allocation rules, borrowed from NVIDIA's blog~\cite[Figure 1]{mig_blog}.
A legal A100 partition picks instances top-down.
If an instance is chosen (for example, the shaded \fourslice),
then all the instances beneath it are unavailable.
As an example, the shaded boxes represents a legal partition with
  a 4/7, a 2/7, and a \oneslice.
}
\label{fig:migalloc}
  \vspace{-3ex}
\end{figure}
 
Figure~\ref{fig:migalloc} depicts the basic MIG allocation rules.
But, there are several exceptions.
For example, ``3/7 + 4/7'' is permitted in the figure but prohibited in practice
and ``3/7 + 3/7'' is possible but not shown in the figure.
In total, there are 18 distinct legal instance combinations in one A100 GPU
(see the full list in NVIDIA's document~\cite{mig_doc}).

Note that the challenge of
allocating a larger-than-1/7 instance is different from
allocating a chunk of consecutive resources, like memory. If there are $n$ free pages,
a memory allocator can always allocate a chunk of consecutive $n$ pages
by a series of memory copies.
Nevertheless, even a GPU has three available slices,
it cannot allocate a \threeslice if a \fourslice has been allocated,
which is a hard-coded rule.

\subsection{A study of serving performance with MIG}
\label{s:migperf}
\label{s:study}

\begin{figure*}[htb]
  \begin{subfigure}{0.49\textwidth}
    \vspace{1.5ex}
    \includegraphics[width=0.49\linewidth]{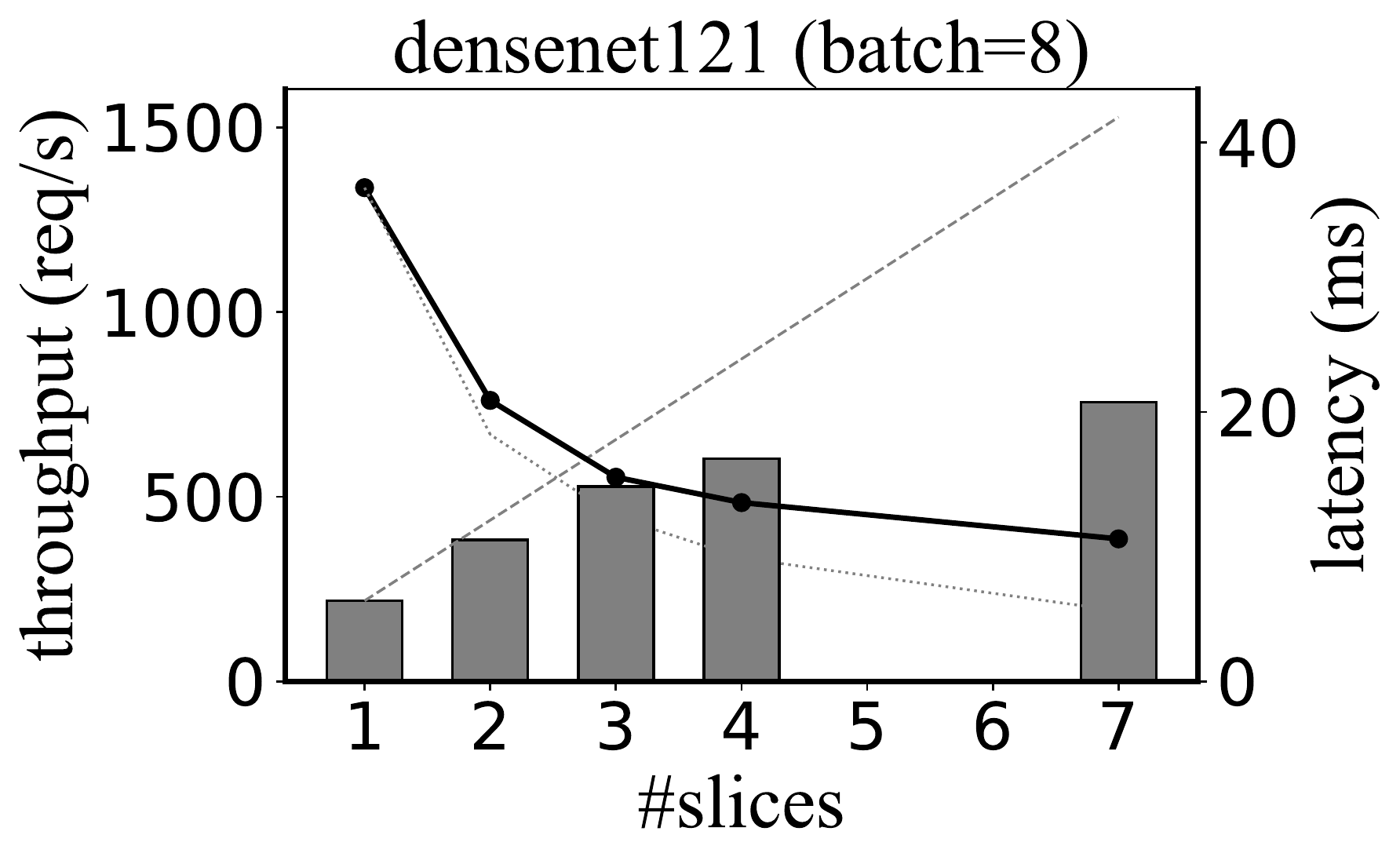}
    \hfill
    \includegraphics[width=0.49\linewidth]{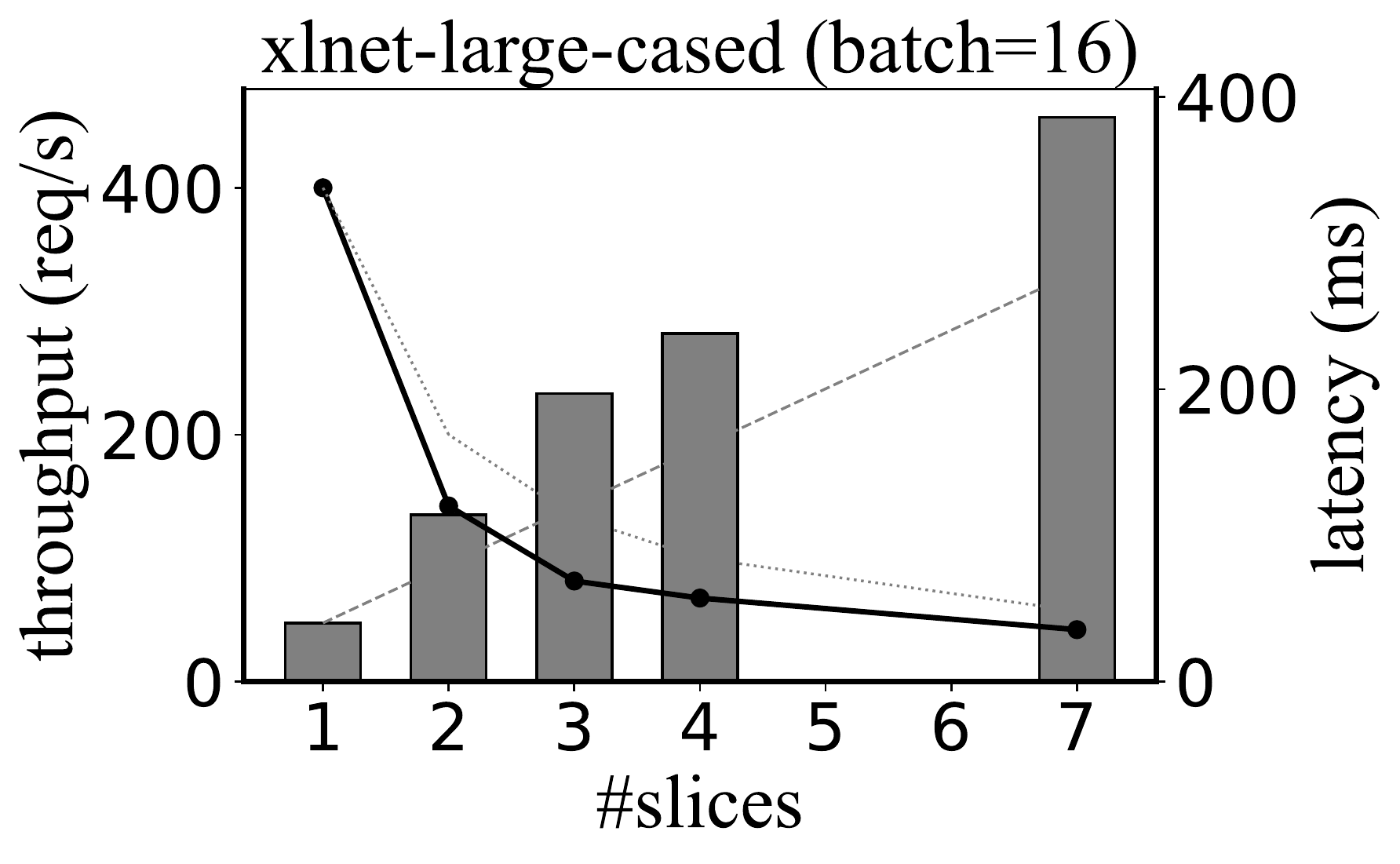}
    \vspace{2ex}
    \caption{Model throughputs and latencies of instances of different sizes.
    The x-axis is the instance sizes, where ``1'' represents a \oneslice, and so on.
    Bars represent throughputs of the corresponding instances; the solid line represents the 90\%-tile latencies.
    The dashed and dotted lines indicate the throughputs and latencies if the
    model's inference performance grows linearly.}
    \label{fig:instances2}
  \end{subfigure}
  \hfill
  \begin{subfigure}{0.49\textwidth}%
    \includegraphics[width=0.48\linewidth]{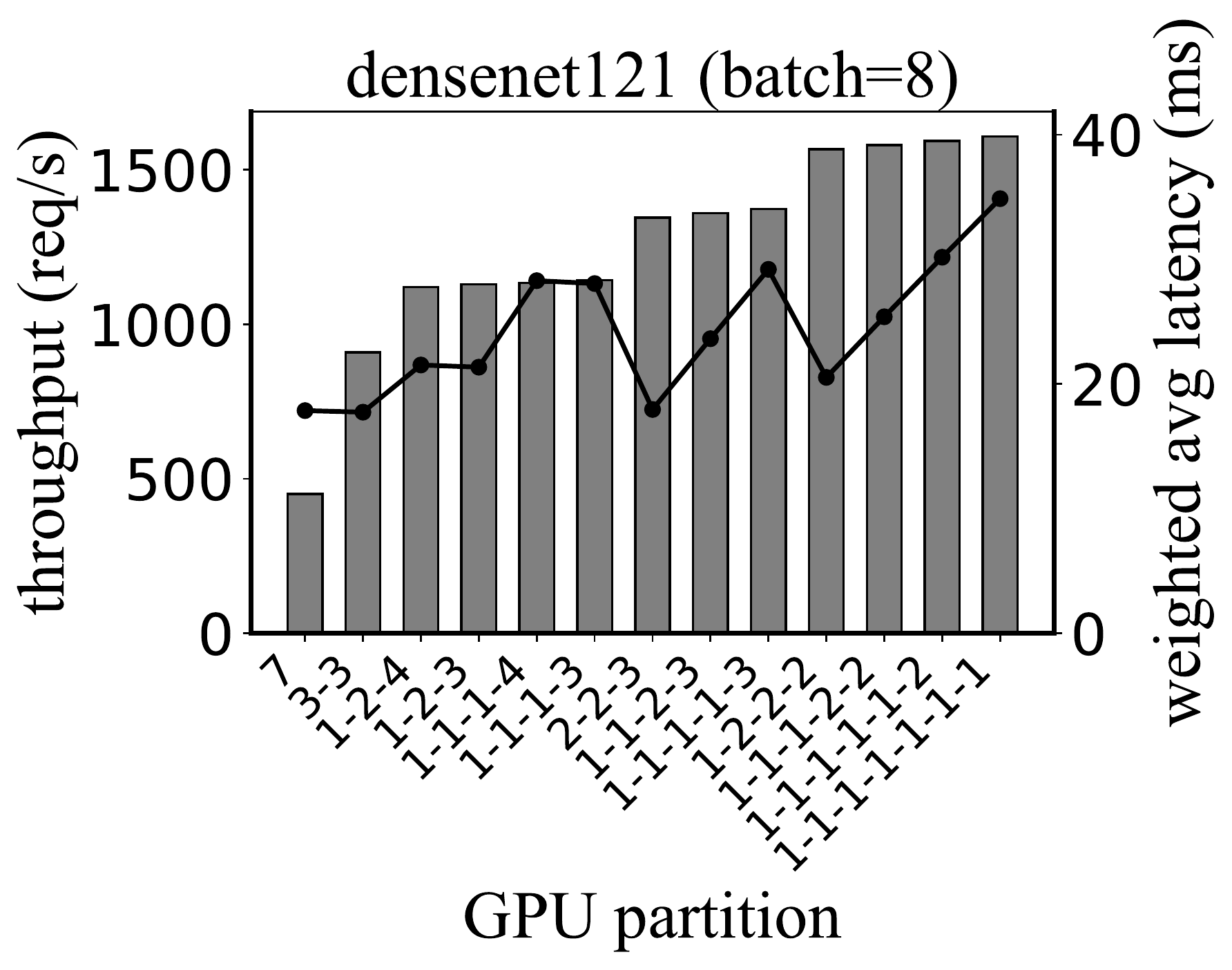}
    \hfill
    \includegraphics[width=0.48\linewidth]{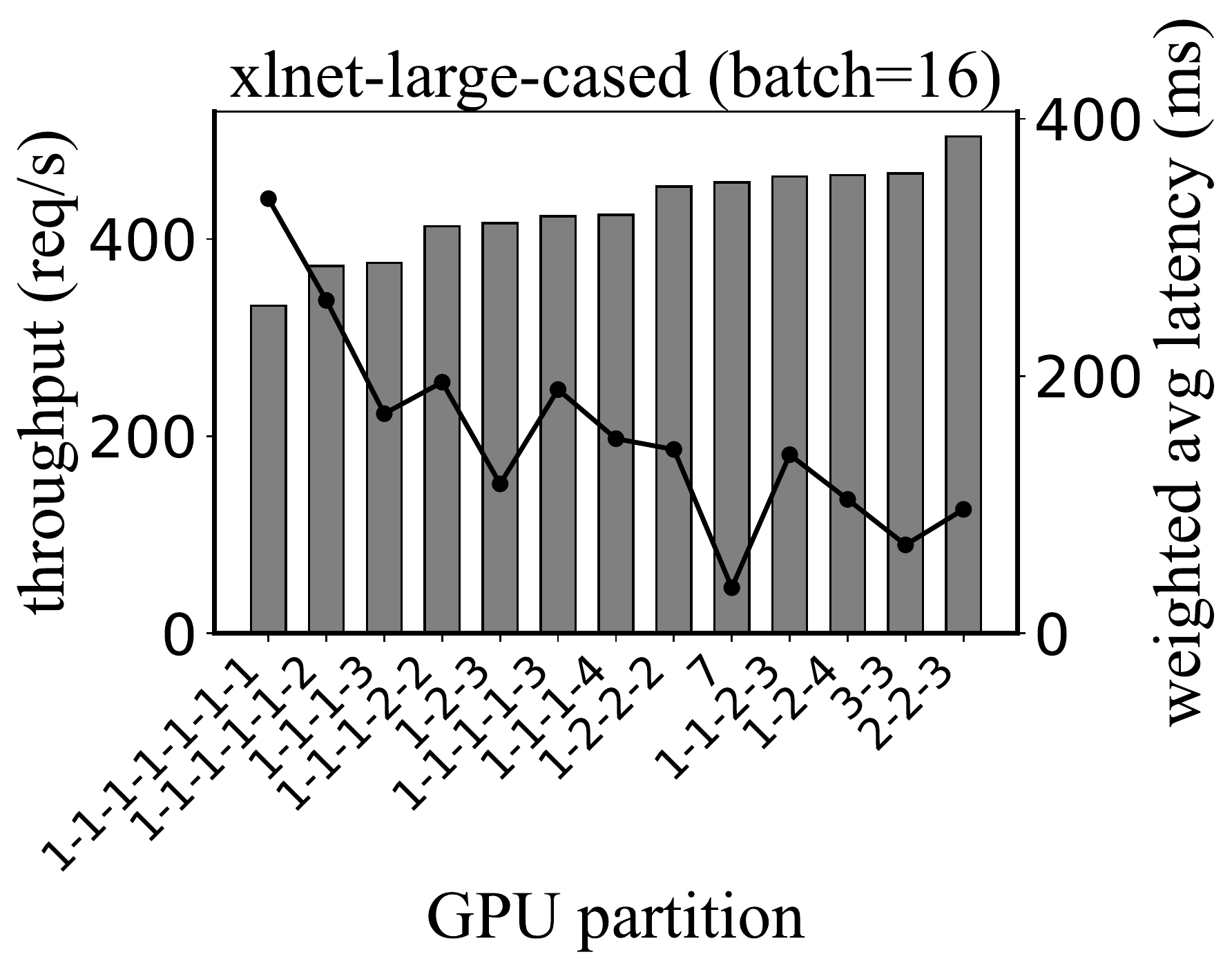}
    \caption{Model throughputs and latencies for different GPU partitions.
    The x-axis is GPU partitions, sorted by throughputs.
    In x-ticks, digits represent instance sizes, for example, ``3-3''
    represents two \threeslice{}s.
    Bars represent throughputs and lines represent the weighted latencies
    which is the averages latencies from different instances
    weighted by their throughputs.
    }
    \label{fig:gpus2}
  \end{subfigure}\hfill
\caption{Throughputs and 90\%-tile latencies for two PyTorch models,
  \modelone and \modeltwo,
  on different sized instances (Figure~\ref{fig:instances2})
  and different GPU partitions (Figure~\ref{fig:gpus2}).
  }
\label{fig:tputlats2}
\end{figure*}
 
To understand the serving performance on different sized instances,
we conduct a study of 49 trained DNNs,
in which 24 models are from PyTorch Hub~\cite{pytroch_hub}
and 25 are from TensorFlow Hub~\cite{tensorflow_hub}
(see Appendix~\ref{appx:study} for all the models).
The models are with precision of FP32.
We run models on 1/7, 2/7, 3/7, 4/7, and 7/7 instances each for 5 minutes, and collect their throughputs and 90\%-tile latencies.
Figure~\ref{fig:tputlats2} shows the results of two
PyTorch models---\modelone and \modeltwo---which represent
two categories of models.
We use them as
illustrative examples below.

By analyzing the throughput and latency trends in Figure~\ref{fig:tputlats2}, %
we have three main observations:

\heading{Observation 1 (Figure~\ref{fig:instances2}):}
the growth of inference throughput is non-linear
relative to the increase in resources (i.e., from 1/7 to 7/7 instances).
Some models (like \modelone) has sub-linear throughput growth,
while others (like \modeltwo) have super-linear throughput growth.
Of course, there are models whose throughputs grow
linearly (see examples in Appendix~\ref{appx:study}). %
But the point is, models scale differently, %
hence a unit of resource contributes differently
for different models and instances.

\heading{Observation 2 (Figure~\ref{fig:gpus2}):}
for the same DNN model,
a GPU with different partitions
has diverse performance, in terms of throughput and latency.
As shown in Figure~\ref{fig:gpus2},
with the same resources (an A100 GPU) but different partitions,
throughputs may differ by up to 4$\times$ (for \modelone);
the latencies vary up to 8$\times$ (for \modeltwo).

\heading{Observation 3 (Figure~\ref{fig:instances2}, \ref{fig:gpus2}):}
The performance characteristics of different DNN models are different.
By pairwise comparing the performance of the two models in Figure~\ref{fig:tputlats2},
we see that models have different performance patterns,
and they prefer different GPU partitions.
For example, \modelone prefers small instances,
as \oneslice has the highest per-unit-throughput without sacrificing
too much on the latency---a 20ms latency increase versus an \sevenslice.
On the contrary, \modeltwo should prioritize large instances
because they have higher per-unit-throughput and lower latency
than smaller instances.

\heading{Model performance classification.}
To understand the performance characteristics across models,
we classify models into three categories,
based on their throughput growth trends:
(1) linear models whose throughputs grow linearly with the increase of computational resources,
(2) sub-linear models whose throughputs grow sub-linearly,
and (3) super-linear models whose throughputs grow super-linearly.

We classify a model into the three categories as follows.
For a model $M$, we calculate a per-unit-throughput for the smallest instance
that can run $M$ (usually \oneslice, but sometimes 2/7 or 3/7 instance if $M$ is large).
Then, we calculate the ratio of \sevenslice's throughput and the above per-unit-throughput.
If the ratio is within $[6.5, 7.5]$, we call $M$ a linear model;
if the ratio is smaller than $6.5$, $M$ is a sub-linear model;
or else, $M$ is a super-linear model.

\begin{figure}[t]
  \begin{center}
  \includegraphics[width=0.45\textwidth]{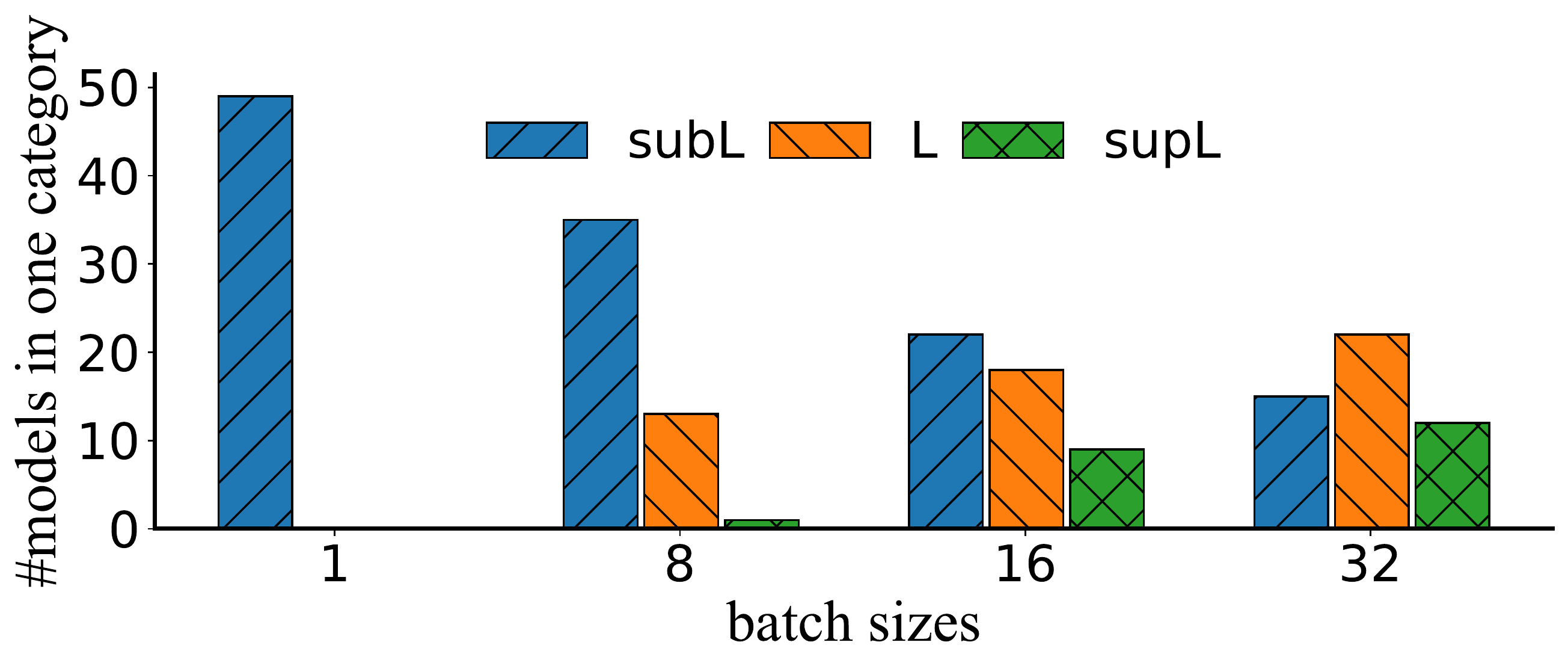}
  \end{center}
  \vspace{-2ex}
  \caption{
  Model classifications.
  ``subL''/``L''/``supL'' indicate sub-linear/linear/super-linear models, respectively.
  }
  \vspace{-3ex}
  \label{fig:modelclassification}
\end{figure}

Figure~\ref{fig:modelclassification} depicts the classification of all
49 models on different batch sizes.
We learn that non-linear models are prevalent,
which account for majority of the cases.
In general, we should assume that a unit of resource contributes
differently in different sized instances for most models.
Another takeaway is that, when the batch size increases,
models are more likely to behave as linear or super-linear.
This is intuitive because the larger the batch, the heavier the computation,
thus models can saturate (or over-occupy) the hardware resources.

Admittedly, our classification is rudimentary.
Comprehensively understanding DNN performance requires further research.
Nevertheless, this basic classification sheds some light
on interpreting DNN model performance with MIG.
Indeed, based on the above observations and classifications,
we invent some heuristics which help the performance of our scheduling
algorithms (\S\ref{s:heuristic}).

\subsection{Strawman approaches: homogeneous partition and static partition}
\label{s:homoserving}

At a high level, serving DNNs with MIG
requires first partitioning GPUs into a collection of instances, %
and then assigning models to instances running as \textit{services}
which respond end user requests.
A straightforward approach is to statically partition the GPUs
and treat the service assignments as a classic scheduling problem.
By whether having heterogeneous instances,
we have two baselines as follows.

First, GPUs are partitioned into homogeneous instances
(either 1/7 or \sevenslice{}s),
then the problem of scheduling DNN services on instances
becomes the \textit{Identical Parallel Machine Scheduling Problem}~\cite{min99genetic}.
Second, GPUs are partitioned to heterogeneous instances
(a mix of multiple instance sizes),
thus the problem reduces to
the problem of scheduling jobs in a heterogeneous cluster,
namely the \textit{Unrelated Parallel Machine Scheduling Problem}~\cite{kim02unrelated,pinedo12scheduling}.

The two baselines are not ideal as they ignore MIG's dynamic reconfigurability.
Our goal is to design a system that automatically partitions (and re-partitions)
GPUs and assigns DNN services to the best suited instances.
It turns out that the general problem we face is a new abstract problem,
which we define formally in the next section.
And, serving DNNs with MIG is a case of this abstract problem.
We will circle back with a rigorous problem statement
in section~\ref{s:realproblem}.

\section{The Reconfigurable Machine Scheduling Problem}
\label{s:prob}

We first defines \prob (short as \textit{\shortprob}) in section~\ref{s:formaldef},
then highlights the differences between \shortprob
and related scheduling problems in section~\ref{s:relsched}.
Finally, section~\ref{s:realproblem} describes in detail the problem that this
paper targets---serving DNNs with MIG.

\subsection{Problem definition}
\label{s:formaldef}

We have a set of jobs and a set of machines.
Each machine can process one job at a time.
Different machines have different processing time for different jobs.
Machines are \textit{reconfigurable}: a set of machines
can be rearranged to a different set of machines,
under some pre-defined \textit{\rules} (defined below).
And the goal is to find a sequence of scheduling
and reconfiguration operations that minimizes (or maximizes)
some given objective, for example,
minimizing cost subject to \slos~\cite{narayanan20heterogeneity}.

Formally, the problem is defined as follows.
There is a set of jobs $\mathcal{J}$ and a set of initial machines $\mathcal{M}_0$,
where $\mathcal{M}_0 \subseteq \mathcal{U}_\mathcal{M}$ and
$\mathcal{U}_\mathcal{M}$ is the universe of all possible machines.
The processing time of job $j$ on machine $i$ is denoted as $p_{ij}$.
We assume that all jobs are known ahead of time and machines do not fail.

A \textit{reconfiguration operation} (\opreconf)
replaces some machines (say $mset$) in the current machines (denoted as $\mathcal{M}_k$)
by another set of machines (say $mset'$).
The \opreconf does not affect jobs running on machines other than $mset$,
which is $\{ m_j | m_j \in \mathcal{M}_k \land m_j \not\in mset\}$.
We call the available machines after a reconfiguration $\mathcal{M}_{k+1}$
and $\mathcal{M}_{k+1} = (\mathcal{M}_k \setminus mset) \cup mset'$
because $mset'$ replaces $mset$.

\textit{\CF{\rules}} (denoted as \rulereconf) specify
whether an \opreconf is legal.
For example, whether two machines can be merged into a larger machine
(an analogy to the rule of merging two consecutive \oneslice{}s, Figure~\ref{fig:migalloc}).
Note that the contents of \rulereconf are specific to problems
and are not part of the \shortprob definition.
As an example, for serving DNNs with MIG, \rules follow MIG partition rules.
The definition of \rulereconf is:
\begin{align*}
rule_{reconf}&(mset, mset', \mathcal{M}_k) \ \to\ Boolean, \\
             &\textrm{where } mset \subseteq \mathcal{M}_k\ \land\ mset, mset' \subseteq \mathcal{U}_\mathcal{M}
\end{align*}
We say a reconfiguration operation $op_{reconf}(mset, mset')$ is legal,
if and only if $rule_{reconf}(mset, mset', \mathcal{M}_k)$ returns $True$.

\heading{Fitting into the scheduling framework.}
\shortprob can be described by the classic scheduling framework~\cite{pinedo12scheduling},
as a triplet $(\ \alpha\  |\  \beta\  |\  \gamma\ )$.
The $\alpha$, $\beta$, and $\gamma$ are three pieces of information
characterizing a scheduling problem:
\begin{myitemize2}

\item $\alpha$ indicates the machine environment. For example, \textit{unrelated
machine in parallel} ($R_m$) is one type of $\alpha$ in which machines
run in parallel and different machines process different jobs at different speeds.

\item $\beta$ describes processing characteristics and constraints, for
example, preemption. %

\item $\gamma$ represents the objective to minimize (or maximize),
for example,
minimizing total cost regarding \slos ($Cost_{min}$).

\end{myitemize2}

\noindent
We see machine reconfigurability
(\opreconf and \rulereconf)
as a member of $\beta$ field, and we denote it as $reconf$.
Thus, \shortprob can be simply read as:

\vspace{-3ex}
\[
 (\ R_m\  |\ reconf\ |\ *\ ).
\]
\vspace{-3ex}

The above asterisk ("$*$") indicates that \shortprob's objectives
are subjective to change for different problems.
For example, with $\gamma = Cost_{min}$,
the problem becomes searching for a series of scheduling and reconfiguration operations
that minimizes the cost while preserving \slos.
This problem $(R_m |reconf| Cost_{min} )$
is the focus of this paper
(detailed description in \S\ref{s:realproblem})

\subsection{Related scheduling problems}
\label{s:relsched}

Scheduling is a broad topic that has been intensively studied.
There are prior scheduling problem variants that consider
reconfiguration in several
forms~\cite{landers01reconfigurable,jozefowska01solving,gorczyca09discrete,xing00parallel,mahmoodjanloo20flexible,azab15modelling,jansen20approximation},
but none of them
fully captures the characteristics of MIG.
We elaborate the most relevant ones below
(see more in \S\ref{s:relwork}).

A recent work that is closely related to our problem (\shortprob) is FJSSP-CDST~\cite{mahmoodjanloo20flexible}
(\textit{Flexible Job Shop Scheduling Problem with machine Configuration-Dependent Setup Times}).
This is a problem combining a classic scheduling problem FJSSP~\cite{pezzella08genetic,pinedo12scheduling}
and a module named RMTs~\cite{landers01reconfigurable} (\textit{Reconfigurable Machine Tools}).
An RMT is a fixed group of machines that can be deployed with different
configurations to serve multiple manufacturing purposes.

\shortprob differs from FJSSP-CDST in the way how reconfigurations behave.
FJSSP-CDST has a basic reconfigurable unit (an RMT)
which contains a fixed group of machines.
During a reconfiguration, all machines in this unit have to stop.
This is a restriction to our (hence MIG's) reconfigurability
because we do not dictate which machines have to be reconfigured at the same time;
for example, an A100 GPU can merge two \oneslice{}s without affecting other instances.

Other related scheduling problems include
DCSP~\cite{gorczyca09discrete,jozefowska01solving} (\textit{Discrete-Continuous Scheduling Problem}) and
UPM~\cite{pinedo12scheduling,kim02unrelated} (\textit{Unrelated Parallel Machine Scheduling Problem}).
The former, DCSP, studies the continuously divisible resources (for example, power),
whereas resources in GPUs are discrete (organized and exposed in instances)
and are constrained in allocation---for example,
allocating a \threeslice %
requires no \fourslice in the same GPU.
For the latter, \shortprob shares the same machine environment ($R_m$) with UPM,
but UPM does not consider machine reconfigurations.

\subsection{A case of \shortprob: serving DNNs with MIG}
\label{s:realproblem}
\label{s:actualproblem}

This paper focuses on a variant of \shortprob---serving DNNs on GPUs with MIG.
In this problem,
machines are GPU instances; %
jobs are DNN services;
different services have different performance on different sized instances
(DNNs' non-linear performance, \S\ref{s:study}).
A set of instances in one GPU can be re-partitioned to another set of instances
(a reconfiguration),
without affecting other running instances on the same GPU.
Our goal is to find the most efficient GPU partitions and service assignments
that minimizes the number of GPUs used.

A reconfiguration is valid when it follows the MIG partition rules (\S\ref{s:migintro}), defined below.
\begin{align*}
rule_{reconf}&(mset, mset', \mathcal{M}_k) \triangleq \\
             &\ \ \ \ \forall m \in mset \cup mset', \ m \textrm{ is in the same } GPU_i \\
             &\land  \mathcal{M}_k |_{GPU_i} \in \textrm{legal A100 partitions} \\
             &\land  \mathcal{M}_k |_{GPU_i} \setminus mset \cup mset' \in \textrm{legal A100 partitions}
\end{align*}
In the above definition, $mset$ and $mset'$ are GPU instances before and after the reconfiguration.
The reconfiguration succeeds iff all instances in $mset$ and $mset'$ are from the same $GPU_i$,
and the GPU partitions before and after reconfiguration
($\mathcal{M}_k |_{GPU_i}$ and $\mathcal{M}_k |_{GPU_i} \setminus mset \cup mset'$)
are legal A100 partitions.

One characteristic of serving DNNs is that jobs (services) are
``long-running'': they do not finish until a shutdown or an update.
This is a simplification compared to the general \shortprob
because it spares the decisions on job scheduling timing.
In particular, we do not have to consider
when to schedule a job (service) because they all need to be deployed in the beginning
and are long-running.

Serving DNNs with MIG is an NP-hard problem
because an NP-hard problem, \textit{Cutting Stock Problem}~\cite{np_prob}, reduces to it.
The cutting stock problem studies how to cut standard-sized paper rolls
into certain numbers of specified-sized pieces while minimizing wasted material.
This problem can reduce to our problem by
treating the paper rolls as GPUs,
specified-sized pieces as different sized instances for services,
and the required piece numbers as \slos.
If one can find the minimum GPUs for our problem,
we know the minimum paper rolls for the original problem,
which minimizes the wasted materials.

\section{System overview}
\label{s:sys}

To serve DNNs with MIG efficiently,
we design and implement a system, \sys,
which automatically partitions GPUs and assign services.
This section introduces \sys's design
and its main components: \optimizer and \scheduler.

\begin{figure}[t]
\begin{center}
\includegraphics[width=0.48\textwidth]{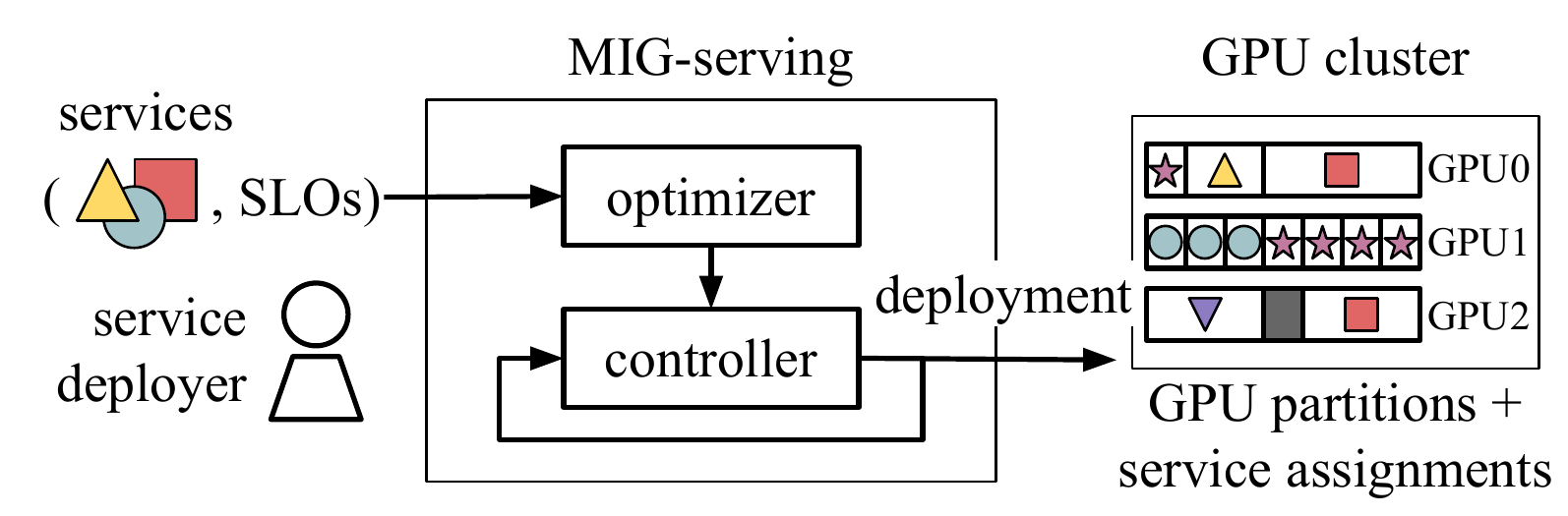}
\end{center}
  \vspace{-3ex}
\caption{\Sys's architecture.}
\label{fig:sysarch}
  \vspace{-3ex}
\end{figure}
 
\heading{Workflow.}
Figure~\ref{fig:sysarch} depicts \sys's architecture.

A \textit{service deployer} specifies what \textit{services} (DNN models) to run
and their \textit{\fullslos} (\slos)
which include required throughputs and expected latencies.

\Sys takes in the services (models) and their \slos as inputs,
and is responsible to produce a \textit{\deployment}---a combination of GPU
partitions and service assignments.
A \deployment is valid if it satisfies \slos:
for each service,
(i) the sum of throughputs from all instances is greater than the required throughput,
and (ii) the 90\%-tile latency of each instance is smaller than what required by \slos.

\Sys then generates a \textit{\plan}
which transfers GPU clusters from the current \deployment
to the newly generated one.
Finally, \sys executes this \plan on GPU clusters.
The entire transition process is transparent to end users; %
they do not experience service interruptions.

\Sys has two main components, \optimizer and \scheduler.
At a high level,
\optimizer designs a valid \deployment for the given \slos,
and \scheduler implements this \deployment transparently.
Next, we briefly introduce these two components.

\heading{\Optimizer.}
\Optimizer tackles the optimization problem of serving DNNs with MIG (\S\ref{s:realproblem}):
finding a valid \deployment that uses as few GPUs as possible.
\Optimizer runs a two-phase algorithm
which blends a heuristic greedy algorithm, a Genetic Algorithm (GA),
and a Monte Carlo Tree Search algorithm (MCTS).
The first phase %
aims at finding a candidate \deployment which is valid but
suboptimal in terms of GPU usage efficiency.
The second phase improves the candidate \deployment
via a combination of custom-designed GA and MCTS.

This two-phase design is crucial in practice
because it balances
two important but conflicting requirements:
(i) getting a valid \deployment quickly
and
(ii) taking full advantage of every single GPU.
The two requirements are at odds because
we need to quickly have at least some plan that satisfies the \slos in case of urgent changes,
but exploring configuration possibilities takes a lot of time.

\Optimizer's first phase runs a fast algorithm (the heuristic greedy algorithm)
in $O(n^2 m)$ where $n$ and $m$ is the number of services and GPUs,
which can produce results in minutes;
whereas the second phase is expensive
and runs continuously and massively in parallel.
Note that the second phase is on-demand.
People can decide how much time and how many computational resources %
they are willing to devote.

\heading{\Scheduler.}
\Scheduler receives two inputs, the new \deployment (from \optimizer)
and the current \deployment on GPU clusters.
\Scheduler's duty is to (i) plan a series of \textit{actions} (called a \plan)
that switch GPUs from the current configurations to the new version,
and (ii) execute the \plan without affecting user experiences.
\ZC{As discussed, use 'controller' instead of scheduler, since scheduler is lower-level. }

To achieve the above goals, \scheduler runs an algorithm, called \textit{\exchangeandcompact}.
At a high level, the algorithm first changes current service instances to
the wanted sized instances while maintaining the required throughputs
during this process,
with the help of extra GPUs; %
it then repartitions GPUs and packs the services into the planned number of GPUs.

\label{s:actions}
During a \deployment transition,
\scheduler has four types of actions:
instance creation, deletion, migration, and GPU repartition.
These actions are implemented in Kubernetes (\S\ref{s:impl}),
and \scheduler issues actions according to the \plan,
and some actions are issued in parallel (\S\ref{s:algotwo}).

\section{\Optimizer algorithm}
\label{s:algoone}

This section describes how \sys solves an optimization problem
of minimizing number of GPUs used while satisfying \slos.
Section~\ref{s:optprocedure} encodes this optimization problem;
section~\ref{s:twophase} depicts the overall algorithm pipeline of \sys's \optimizer;
and section~\ref{s:fastandslow} introduces the two concrete algorithms \optimizer uses.

\subsection{Defining \optimizer procedure}
\label{s:optprocedure}

As mentioned (\S\ref{s:sys}),
\optimizer is obligated to generate valid \deployment{}s that fulfill \slos.
Next, we define this procedure, %
which provides a basic framework for different algorithms.

\Optimizer's inputs are
(1) service performance (throughput and latency)
on different sized GPU instances (1/7--7/7 instances),
and (2) \slos %
which include required throughputs and latencies for each service.
\CF{\optimizer}'s output is a \deployment %
that consists of GPU partitions and
service assignments.

\label{s:def:sstate}
We define \textit{\sstate} for a \deployment %
to represent the \deployment's progress of satisfying \slos.
\CF{\sstate} is a vector of percentage numbers;
each number represents the percentage of a service's current throughput
to the required throughput in \slos.
For example, a \deployment has \sstate of $[0\%, 100\%, \cdots]$
means that the \deployment does not run \serv{0} on any GPUs
while \serv{1} is fully satisfied.

\label{s:def:utility}
For a service running on an instance, we calculate a \textit{utility}
which indicates how much this instance contributes to the service's total
throughput requirement.
For example, if \serv{0} requires $1000$ req/s %
and a \oneslice has a throughput of $50$ req/s for \serv{0},
then we say \serv{0} on \oneslice has a \utility
of $[+5\%, 0\%, \cdots]$ (we use ``$+$'' to distinguish \utility from \sstate).
With the \utilities of all services,
we can calculate the \utility for a GPU
by adding up the \utilities of all instances in this GPU:
for the same example of \serv{0},
if a GPU has seven \oneslice{}s running \serv{0},
it has a \utility of $[+35\%, 0\%, \cdots]$
($35\% = 7 \times 5\%$).

\label{s:configspace}
Note that the \utility space for all possible \gconfig{}s is enormous.
A loose upper bound is $O(n^7)$ where $n$ is the number of services,
because a GPU has at most $7$ instances and each instance can run one of $n$ services.
Of course, the actual size is much smaller than this bound.
Nevertheless, it is still huge; the number of possible \gconfigs (\utilities)
is 157.8k and 234.7k when $n$ is 12 and 13, respectively.

Finally, we define an \optimizer procedure as follows.
Given (i) \utilities for all service on all sized instances and (ii) \sstate,
an \optimizer procedure should produce a set of \gconfigs, such that the sum of all \gconfig \utilities and the \sstate
must be greater than or equal to
$[100\%, 100\%, ...]$
(with respect to vector comparison).
Note that the given \sstate is not necessarily all zeros.

\subsection{Two-phase algorithm and GA}
\label{s:twophase}

\begin{figure}[t]
\begin{center}
\includegraphics[width=0.35\textwidth]{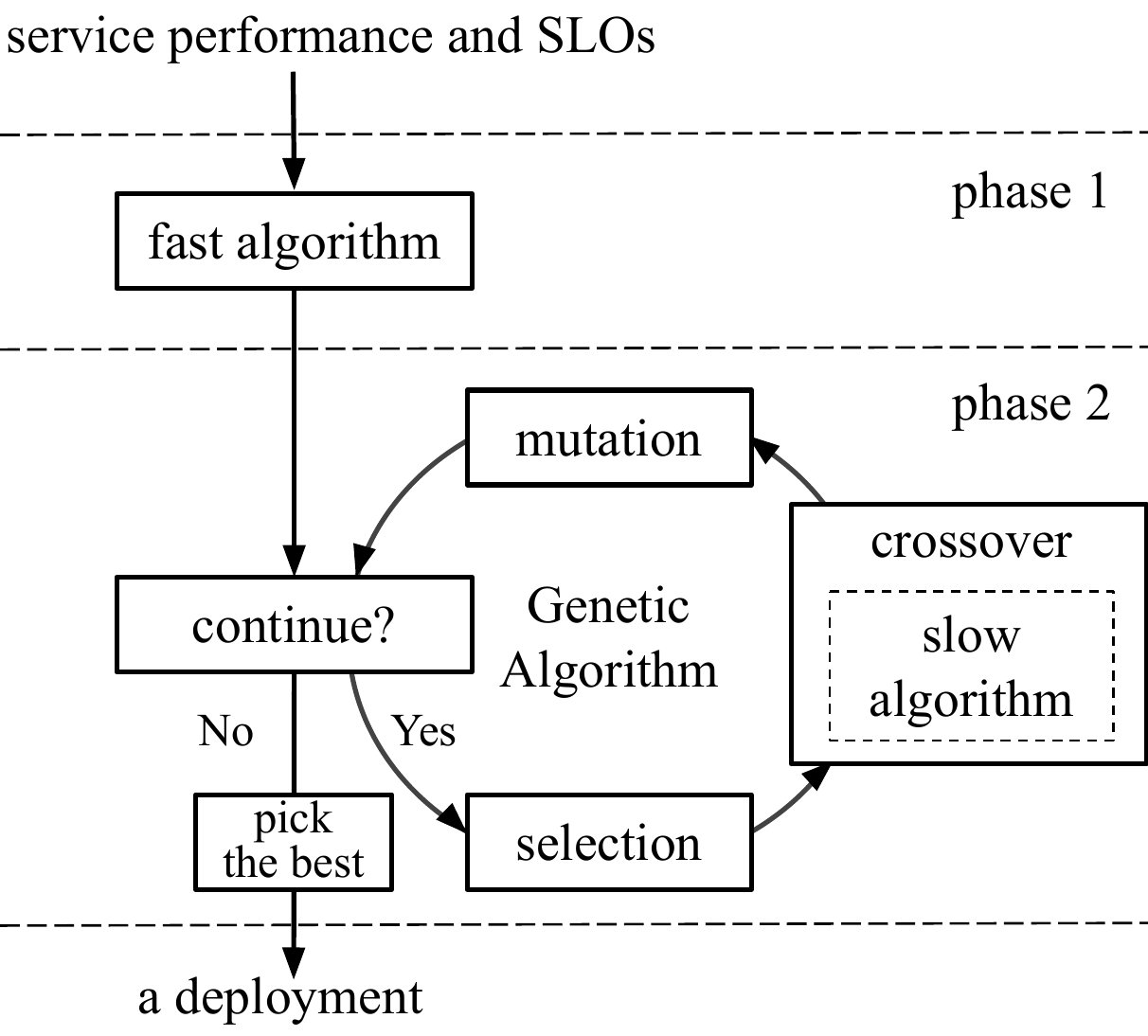}
\end{center}
  \vspace{-3ex}
\caption{\Sys's two-phase algorithm. The fast algorithm and the slow algorithm
are optimizer procedures (\S\ref{s:optprocedure}).}
\label{fig:algofastandslow}
  \vspace{-3ex}
\end{figure}
 
\Sys's \optimizer runs a two-phase algorithm,
which is outlined in Figure~\ref{fig:algofastandslow}.
It has two ``template'' algorithms,
namely the \textit{fast algorithm} and the \textit{slow algorithm}.
Both template algorithms
must be an \optimizer procedure (\S\ref{s:optprocedure}),
and the fast algorithm is supposed to run (relatively speaking) fast.
The two algorithms are connected by the two-phase design
and a custom-designed Genetic Algorithm (GA for short).
In the rest of the section,
we first introduce the properties of the fast and the slow algorithms,
and then describe the custom GA and its two main pieces, \textit{crossover}
and \textit{mutation}. %

\heading{Fast and slow algorithms.}
In our design, we require the fast algorithm
(i) to be a legal \optimizer procedure,
and (ii) running fast---the algorithm's time complexity
must be polynomial
with respect to the number of services and GPUs.
In practice, we require the algorithm to finish in minutes.

For the slow algorithm, we only require it to be a legal \optimizer procedure.
Nevertheless, we expect the slow algorithm to discover better
solutions than the fast algorithm (hopefully in high probability).
This expectation ought to be possible as the slow algorithm is given more time budgets.

In \sys, we use a heuristic greedy algorithm as the fast algorithm,
and Monte Carlo Tree Search (MCTS) as the slow algorithm (\S\ref{s:slowalgo}).
Of course, they are changeable.

\heading{GA overview.}
GA is a heuristic algorithm framework inspired by nature gene evolution.
We tailor GA to our context:
a chromosome is a \deployment,
and genes are \gconfigs. %
To evolve, a chromosome (\deployment{}) conducts crossovers and mutations.
A crossover erases some \gconfigs in a \deployment
and fills in with new \gconfigs generated by the slow algorithm.
A mutation swaps services running on instances in a \deployment.

GA runs in rounds.
In each round, we select the best \deployments in the last round,
and let them attend the coming crossovers and mutations.
GA stops when time out, or the best \deployment stops
improving in the past ten rounds.
Note that GA keeps the original \deployments in each round's comparison,
so that the best candidate only improves over time.

\heading{Crossover.}
A crossover applies to a (valid) \deployment,
which contains two steps.
First, we \textit{randomly} erase some \gconfigs,
which decreases the overall throughputs and makes some services unsatisfied.
As a result, we have \sstate that are not all-$100\%$.
Second, we run the slow algorithm against the current \sstate
and get a \deployment that makes up for the previously erased GPUs.
The figure below is an illustrative example. %
Each rectangle represents an instance; and different colored symbols (e.g., stars, triangles, squares)
represent different services.

\begin{figure}[H]
\vspace{-1ex}
\centering
\includegraphics[width=0.45\textwidth]{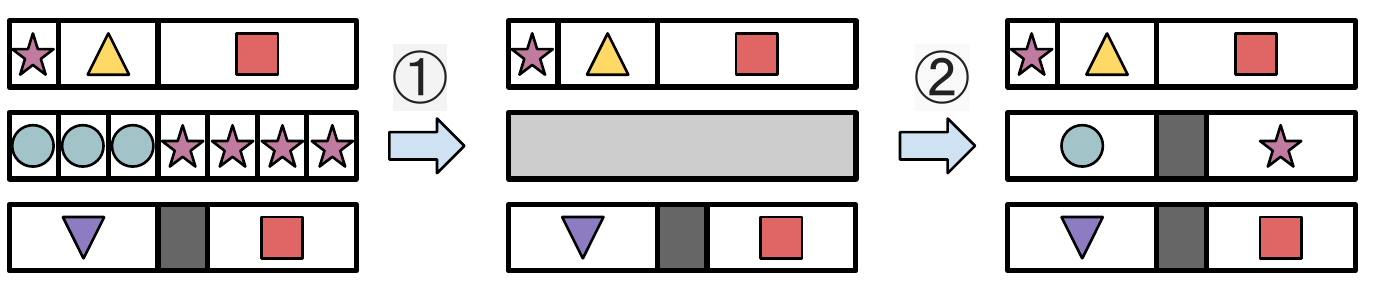}
\vspace{-2ex}
\end{figure}

The insights behind this crossover approach are twofold.
First, a crossover mixes solutions from the slow algorithm and the fast algorithm,
thereby providing diversity.
Second, the problem size of crossovers is much smaller than the original one,
hence the slow algorithm can finish in reasonable time.

\heading{Mutation.}
Mutation is based on an observation that DNN inference does not have affinity requirements
(different from DNN training);
that is, instances of the same size are identical for inference.
A mutation \textit{randomly} picks some instance pairs;
each pair contains two instances that are the same in size but run different services.
The Mutation then swaps the services in each pair.
The figure below depicts this process.
\begin{figure}[H]
\vspace{-1ex}
\centering
\includegraphics[width=0.45\textwidth]{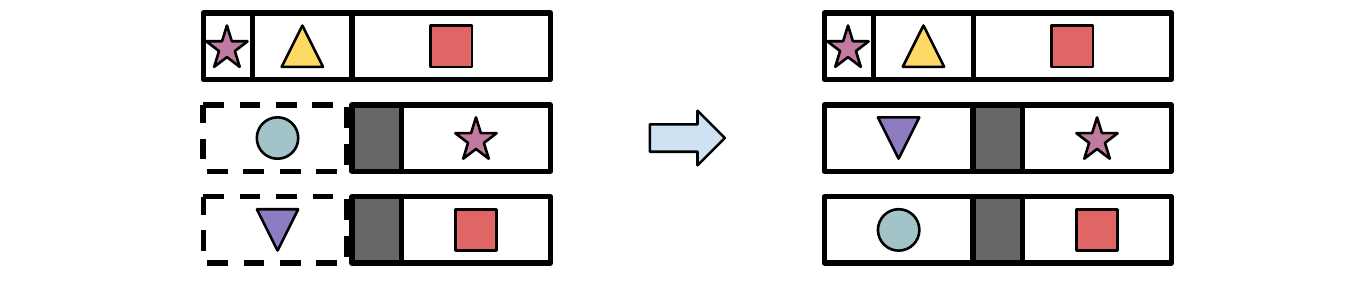}
\vspace{-2ex}
\end{figure}

The idea of mutation is to explore different combinations of services on one GPU.
Mutations themselves do not improve \deployments.
But they create diverse service mixing on GPUs
which helps crossovers explore combining different services.

\subsection{Algorithm, fast and slow}
\label{s:fastandslow}

In this section, we introduce the fast and the slow algorithms used in \sys.
The two algorithms are chosen from a pool of candidate algorithms. It is possible that we may find better algorithms in the future.
\Sys is designed to be able to switch algorithms easily.

\heading{Fast algorithm: heuristic score and greedy algorithm.}
\label{s:fastalgo}
\label{s:score}
\label{s:heuristic}
\Sys develops a greedy algorithm as the fast algorithm,
which chooses the ``best'' \gconfigs according to a \textit{heuristic score}.
This score represents how well a \gconfig
serves the current \textit{service requirements},
that is the complementary to the current \sstate
(namely, $\mathbbm{1} - $ \sstate,
where $\mathbbm{1}$ is an all-1 vector).

The greedy algorithm works as follows.
First, it
ranks the known \gconfigs by their scores
and picks the one with the highest score.
The algorithm then updates the \sstate and repeats the above step until
all service throughputs reach or exceed \slos.
(Appendix~\ref{appx:fastalgo} describes this heuristic greedy algorithm in detail.)

The heuristic score of a \gconfig
is based on two factors: the current \sstate (a vector of percentages, \S\ref{s:def:sstate})
and the \gconfig's \utility (a vector of percentages, \S\ref{s:def:utility}). Below is the score's definition,
where $c_i$ and $u_i$ are the $i_{th}$ number in the \sstate and the \utility, respectively;
$\mathbbm{1}$ is an all-1 vector;
$n$ is the number of services. \begin{align*}
\textrm{score(config)} &= \sum (\mathbbm{1} - \textrm{\sstate}) \odot \textrm{config's \utility} \\
               &= \sum_0^{n-1} (1 - c_i) \times u_i
\end{align*}

The idea behind the score is to balance
a GPU's overall throughputs and the current service needs.
On the one hand, higher throughputs are likely to have higher scores.
On the other hand, the \gconfigs which contribute to services with low \sstate
are likely to have higher scores.
For example, if a configuration $config_a$ has higher throughputs than $config_b$,
then $config_a$ has a higher score. However, if all services that $config_a$ contributes to
are fully satisfied, then the throughputs don't count and $config_a$'s score is $0$.

\heading{Slow algorithm: MCTS.}
\label{s:slowalgo}
\Sys tailors the Monte Carlo Tree Search (MCTS) as the slow algorithm.
We choose MCTS because the problem of allocating MIGs can be naturally encoded to a tree search problem;
hence MCTS, as a well-studied tree search algorithm, is a good candidate.

\begin{figure}[t]
\begin{center}
\includegraphics[width=0.35\textwidth]{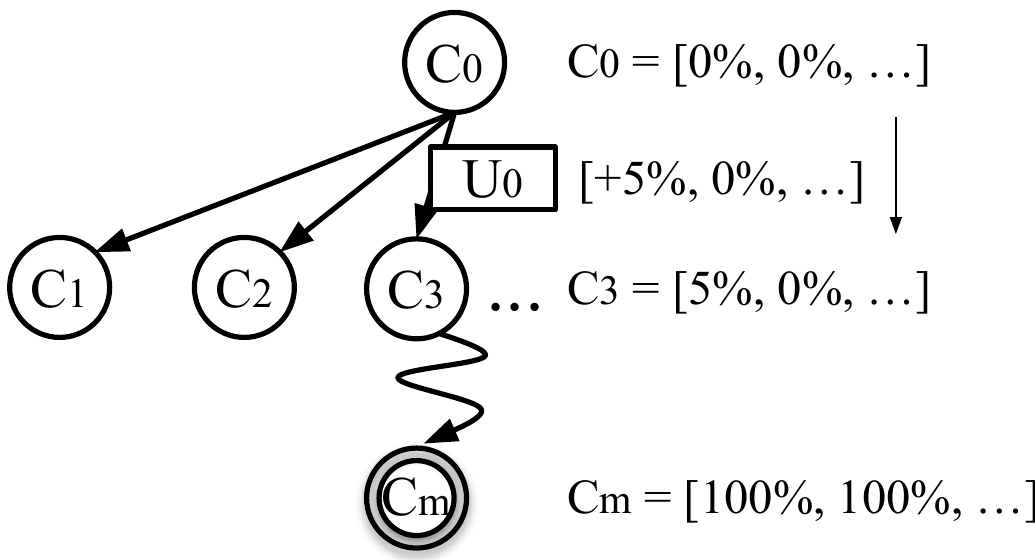}
\end{center}
\vspace{-3ex}
\caption{An example tree for the problem of serving DNNs with MIG.
The nodes (denoted as $C_i$) represent \sstate, and $C_m$ is a leaf node.
Edges (denoted as $U_j$) represent \utilities of \gconfigs.}
\label{fig:mcts}
\vspace{-3ex}
\end{figure}
 
Figure~\ref{fig:mcts} depicts an example of the tree search problem.
Nodes represents \sstate.
Edges represents \gconfig's \utilities.
A transition from a parent node to a child represents the \sys
picking a \gconfig (indicated by the edge).
Leaf nodes are nodes whose
\sstate are all-100\% (or larger than 100\%),
meaning that all services are satisfied.
The goal is to find the shortest path from the tree root (the initial \sstate)
to a leaf node.
This shortest path is the \deployment that uses the minimal number of GPUs (the
length of the path).

MCTS is designed to search heuristically and find a good solution.
However, vanilla MCTS doesn't work in %
our problem, 
for two reasons.
First, each node has too many children: %
the number of children equals the number of edges, which equals the number of \gconfigs.
As mentioned earlier (\S\ref{s:configspace}), the configuration space is huge.
Second, the classic MCTS estimation procedure (also known as simulation, playout, or rollout)
is slow and inaccurate.
The original estimation is to find a random path to some leaf node
which is used to estimate the depth of a subtree.
However, our problem requires an estimation of the \textit{shortest} path
instead of a random path,
which leads to an extremely inaccurate estimation.

To address the above two problems,
\sys customizes MCTS by
(i) cutting the children of each node into the nodes with the top-K heuristic scores
(K=10 by default)
and (ii) using a fast-and-accurate estimation
via memoization and randomization.
The details of the custom MCTS are described in Appendix~\ref{appx:mcts}.

\section{\CF{\scheduler} algorithm}
\label{s:algotwo}

\Sys targets the real-world serving environment,
in which \slos change from time to time, for example, a shift from daytime to night. In addition, services get updated as well---adding/removing services and upgrading DNN models.
In the above cases, \sys needs to recalculate \deployments to adapt to the new requirements,
and transfer GPU clusters from the old \deployment to the new one.
We call this process a \textit{\deployment transition}.

In \sys, \scheduler is responsible for designing and implementing \deployment
transitions (see Figure~\ref{fig:sysarch}).
A straightforward transition approach is to shut down all services,
repartition underlying GPUs, and then reboot the services.
Of course, this method is unacceptable in practice
because services are unavailable
during transitions.

The goal of \scheduler is to finish a transition
without interrupting user experiences
and finish it quickly.
\CF{\scheduler} introduces an algorithm, \textit{\algotwo}, which 
uses two phases to achieve the aforementioned goals.

\heading{Exchange phase.}
Two \deployments differ in two factors: instance sizes for services and GPU partitions.
Exchange phase addresses the first factor by creating and deleting different sized instances.
First, \scheduler calculates the instance differences between the old and the new \deployments for each service.
We denote the difference as $\Delta_i$ for \serv{i},
which contains the ``diff'' of \serv{i}'s instances.
For example, a $\Delta_i = [+4/7, -2/7]$ means that \serv{i} requires
a new \fourslice and has an unneeded \twoslice.

For each service, \scheduler pairs every new instance (for example, ``$+4/7$'')
with some unneeded instances (for example, ``$-2/7$'') \ZC{Is it guaranteed?} such that the throughputs
of the new instance is equal to or larger than the unneeded instances\ZC{NOTE: larger throughput
is achieved at the cost of extra GPUs which are expensive}.
Note that pairing an unneeded instance which has larger throughputs is not allowed
(for example, pairing ``$+1/7$'' and ``$-7/7$'')
because that may fail providing the adequate throughputs hence affecting user experiences.
Finally, \scheduler has a set of new-unneeded instance pairs and a list
of unneeded instances that cannot pair with any new instances.

\CF{\scheduler} executes each new-unneeded instance pair
by creating the new instance first (using extra GPUs if needed)
and then deleting the unneeded instances.
After finishing all pairs, \scheduler deletes instances in the unneeded list.
During the entire process,
\scheduler guarantees that services
have enough capacity to handle the expected volume of user requests.

\heading{Compact phase.}
After the exchange phase, all services have the wanted sized instances in the new \deployment.
But the GPU partitions haven't been changed yet,
and \scheduler uses more GPUs (i.e., extra GPUs)
than expected because of GPU instance fragmentation.
In the compact phase, \scheduler defragments GPUs by repartitioning GPUs and
migrating instances.

At the beginning of this phase,
\scheduler creates a list of GPUs which are not fully occupied (having unused instances), denoted as $S$.
If $S$ is not empty, \scheduler picks a $GPU_i \in S$,
and gathers some running instances from other GPUs in $S$
such that these instances together can fully occupy $GPU_i$;
\scheduler repartitions $GPU_i$ (if needed),
migrates these instances to $GPU_i$,
and removes $GPU_i$ from $S$ (the $GPU_i$ is now fully utilized).
\ZC{I found the above para a little bit confusing on how $GPU_i$ is repartitioned and instances being migrated to. Maybe change the order of sentences?}
\CF{\scheduler} continues the above process until achieving the new \deployment.

\label{s:migrationopt}
\heading{Optimizations.}
\CF{\scheduler} adopts several optimizations. We list two below.
First, \scheduler is locality-aware---it prioritizes \textit{local} instance
migrations over \textit{cross-machine} migrations. %
In our GPU clusters, each machine has 8 A100 GPUs;
migrating instances within a machine is much cheaper than migrating across machines.
Second, actions can run in parallel if the affected GPUs are separate.
\CF{\scheduler} analyzes the dependencies between actions and executes the
non-conflicting ones simultaneously.

Note that the \algotwo algorithm can happen in different granularities,
depending on how many extra GPUs available.
If there are many, \scheduler can run \algotwo once for all services.
However, if only few extra GPUs are available, \scheduler will run \algotwo in multiple rounds;
in each round, \scheduler only targets a small number of services.
\section{Implementation}
\label{s:impl}

\Sys is implemented in Python and Kubernetes (k8s). %
Figure~\ref{fig:impl} lists the components of \sys implementation.
For \optimizer, we implement the \optimizer procedure (\S\ref{s:optprocedure})
as an abstract class that the fast and the slow algorithms extend.
\Sys can easily switch to other algorithms by implementing them
under the same abstract class.

We implement \scheduler by extending k8s controller~\cite{k8s_controller}.
\Sys's actions---instance creation, deletion, migration, and GPU
partition---are wrappers of k8s operations.
For example, a remote instance migration from machine $A$ to $B$ is a sequence of operations:
creating an instance on machine $B$,
checking if the instance on $B$ is successfully launched,
and deleting the instance on machine $A$.

\Sys always chooses the largest batch sizes possible, as far as the
inference latency is smaller than what required by \slos.
This may result in a service with different batch sizes for different instances.
\Sys relies on load balancing systems to dispatch user requests accordingly.

\begin{figure}
\footnotesize
\centering
\begin{tabular*}{0.8\columnwidth}{@{\extracolsep{\fill}} l  l }
\toprule
    \textbf{\Sys component}     &  \textbf{LOC} \\
\midrule
\Optimizer & \\
  \hspace{2ex} data structures &   2674 \\
  \hspace{2ex} fast algorithm (greedy) &   135 \\
  \hspace{2ex} slow algorithm (MCTS) &   398 \\
  \hspace{2ex} Genetic Algorithm (GA) &   330 \\
\Scheduler &\\
\hspace{2ex} k8s controller & 1318 \\
\hspace{2ex} \algotwo algorithm & 319 \\
\bottomrule
\end{tabular*}
\caption{Components of \sys implementation.}
\label{fig:impl}
\vspace{-3ex}
\end{figure}
\section{Experimental evaluation}
\label{s:eval}

In this section, we answer three questions:

\begin{myitemize2}

 \item How many GPUs does \sys save, versus baselines? %

 \item How long does \sys take to complete a \deployment transition,
      and what components cost the most?

 \item Do \sys's deployments meet \slos in practice?

\end{myitemize2}

\heading{Baselines and workloads.}
We have three baselines with static GPU partitions:
(1) \gpuoneseven, partitioning GPUs into \oneslice{}s,
(2) \gpusevenseven, using A100 GPUs as-is,
and (3) A100-MIX, partitioning all A100 into ``4-2-1''
(a combination of 4/7, 2/7, and 1/7 instances)
and scheduling one service on one GPU.
\gpuoneseven is the most cost-efficient setup as shown in Figure~\ref{fig:motivation}.
\gpusevenseven %
uses GPU the traditional way (ignoring MIG).
A100-MIX represents heterogeneous setup but ignoring the characteristics of
workloads.

To see how well \sys performs compared to the optimal solution (which is
computationally expensive to have),
we calculate an approximate optimality---a lower bound of GPU usage by \emph{ignoring}
MIG's hardware constraints.
In particular, we assume that any combination of instance is possible, and the
minimal number of GPUs can be calculated by always using the
most cost-efficient instance.
Notice that the lower bound is likely impossible to achieve
due to ignoring the hardware constraints.

\vspace{1ex}
\noindent
We uses two sets of workloads in the following experiments:
\begin{myitemize2}

 \item Simulation workloads (requiring hundreds of GPUs):
  we generate four workloads for 24 DNN models.
  In each workload, models' \slo throughputs are generated
  from either normal distributions (for two workloads)
  or lognormal distributions (for the other two workloads).
  The latencies in \slos are set to 100ms, which is
  an acceptable waiting time under most scenarios.

 \item Real-world workloads (requiring up to 16 GPUs):
  we build two real-world workloads for five DDN models running in our GPU clusters.
  We collect 24-hr production throughputs of the five models
  and construct the workloads:
  one workload represents the peak throughputs (called \textit{daytime} workload),
  and the other workload represents the low throughputs (called \textit{night} workload).
  Note that we scale down models' throughputs to fit into
  our testbed which has 24 A100 GPUs, %
  while preserving
  throughputs' relative amounts.

\end{myitemize2}

We run \sys on a 104-core machine with 750G memory running Debian 9.13.
To test real-world workloads,
we have a three-machine GPU cluster with 24 A100 GPU cards
(8-GPU for each machine).
The five DNN models for the real-world workloads are
\texttt{robert-large},
\texttt{bert-base-uncased},
\texttt{albert-large-v2},
\texttt{resnet101},
and \texttt{resnet50}.

\subsection{\Optimizer algorithms}
\label{s:eval:algos}

In this section, we study how many GPUs \sys saves compared to baselines.
The workloads used are the four simulation workloads
generated from normal and lognormal distributions,
denoted as \texttt{normal-1},
\texttt{normal-2},
\texttt{lognormal-1},
and \texttt{lognormal-2}.
We design the four workloads to use several hundreds of GPUs,
representing a median-sized GPU cluster.

\begin{figure}[t]
\begin{center}
\includegraphics[width=0.48\textwidth]{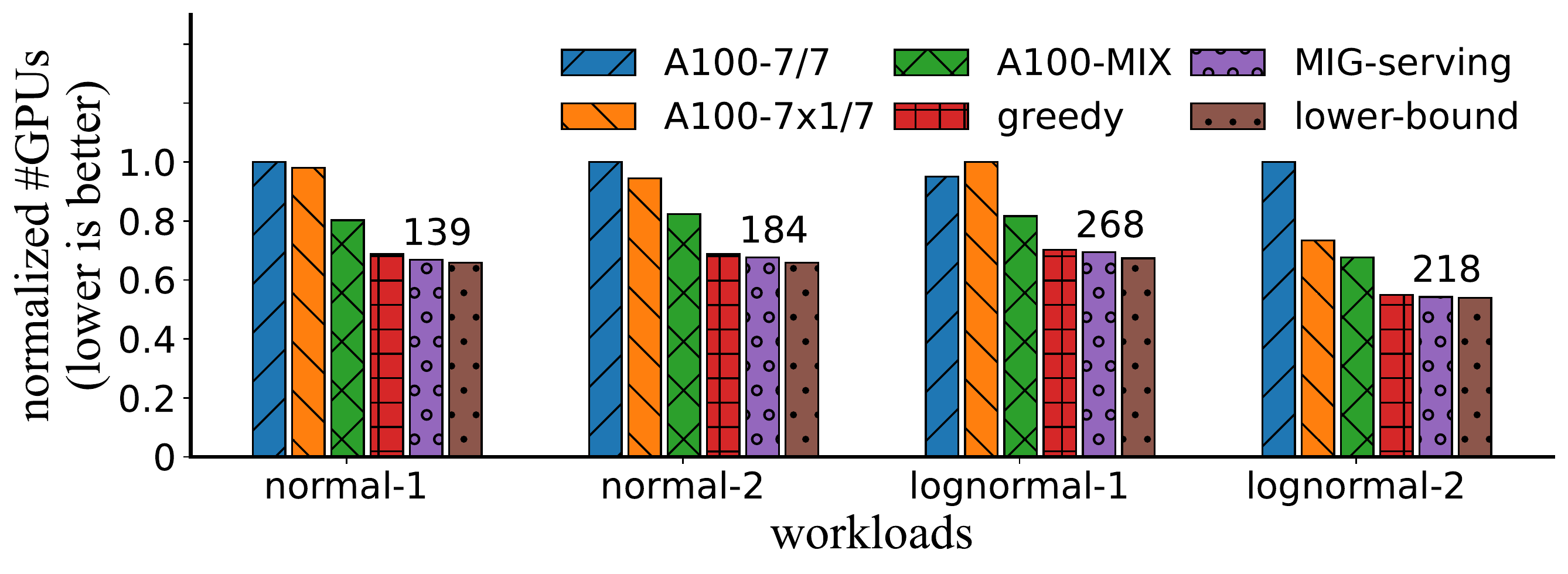}
\end{center}
\vspace{-3ex}
\caption{Number of GPUs used by algorithms for different workloads.
The numbers are normalized clusterwise for each workload.
\Sys's absolute numbers of GPUs used are given.
``\gpusevenseven'', ``\gpuoneseven'', and ``A100-MIX'' are the three baselines;
``greedy'' represents \sys's fast algorithm (\S\ref{s:fastalgo});
``MIG-serving'' represents \optimizer's two-phase algorithm;
``lower-bound'' is the minimal number of GPUs ignoring MIG's hardware constraints.
}
\label{fig:algogpus}
\vspace{-3ex}
\end{figure}
 
\heading{GPU saved.}
We run \sys and the baselines on the simulation workloads
and count the number of GPUs they use.
Figure~\ref{fig:algogpus} shows the results.
\Sys uses fewer GPUs than other baselines.
It saves up to 40\% GPUs compared to \gpusevenseven.
Moreover, \sys is close to the optimal allocation---\sys uses $<$3\% more GPUs
than the GPU lower bound (the ``lower-bound'' in
Figure~\ref{fig:algogpus}).
One thing to clarify is that
\gpuoneseven does not perform as well as in Figure~\ref{fig:motivation}
because solutions now consider latencies, %
hence some models cannot use large batch sizes on \oneslice{}s.

Note that this experiment does not consider the running time of algorithms.
Thus, it is not an entirely fair comparison because
baselines finish in seconds, whereas \sys's fast algorithm finishes in minutes
and the \optimizer's two-phase algorithm finishes in hours
(\sys runs 10 rounds of GA for the four workloads and finishes in 3hr, 5hr, 6.5hr, and 6hr).
But in practice, the service and \slo updates are relatively infrequent (for
example, twice a day), thus we can afford to run \optimizer's two-phase algorithm.
In addition, the \deployment can be reused if models and their \slos are not changed.

\begin{figure}[t]
\begin{center}
\includegraphics[width=0.48\textwidth]{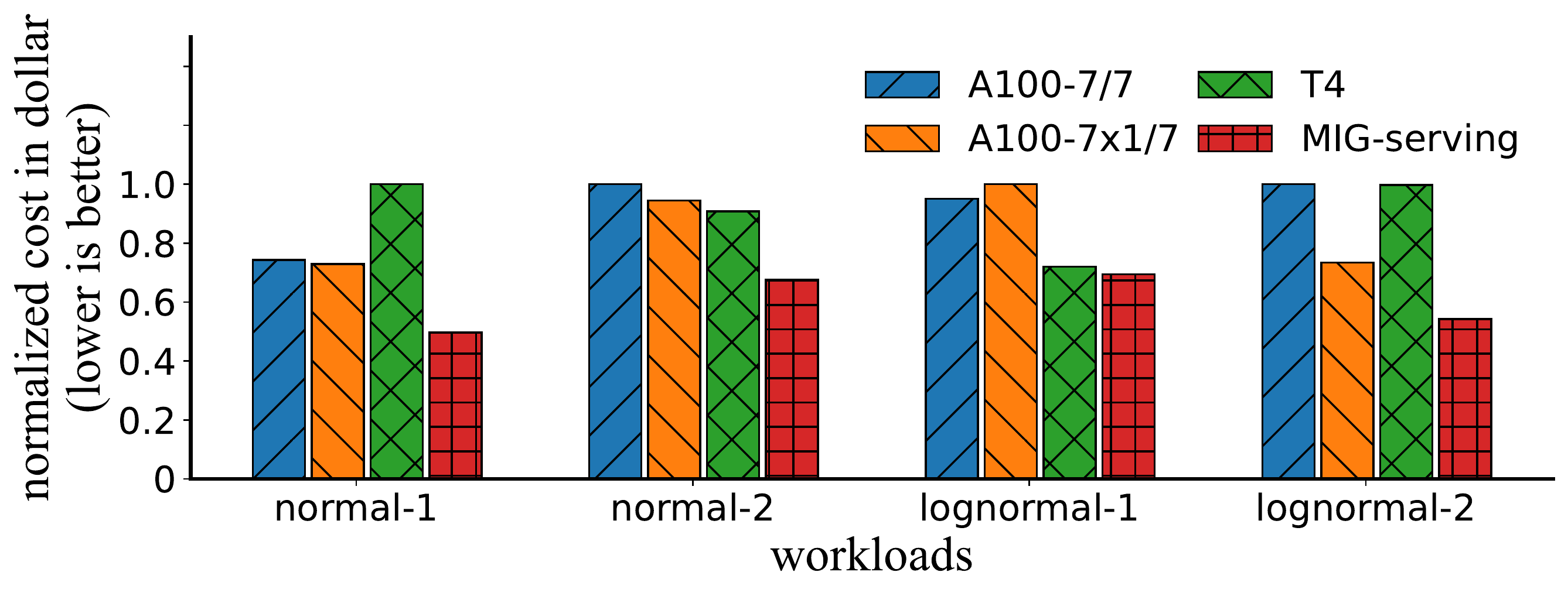}
\end{center}
  \vspace{-3ex}
\caption{Normalized costs for satisfying \slos of different workloads.
``A100-7/7'' and ``A100-7x1/7'' are the two baselines (with A100);
``T4'' represents using T4 GPUs;
``MIG-Serving'' repreesnts the two-phase algorithm with A100.
We use the price from AWS GPU machines~\cite{aws_t4, aws_a100}.
}
\label{fig:t4cost}
\end{figure}
 
\heading{Cost versus T4.}
To compare the cost-efficiency with other GPU types, we evaluate how many T4 GPUs the
simulation workloads need to satisfy their \slos.
We choose T4 because it is the most cost-efficient GPU type for DNN serving
before A100.
Figure~\ref{fig:t4cost} shows the results.
\sys is the most cost-efficient configuration for all workloads.

\begin{figure}[t]
\begin{center}
\includegraphics[width=0.48\textwidth]{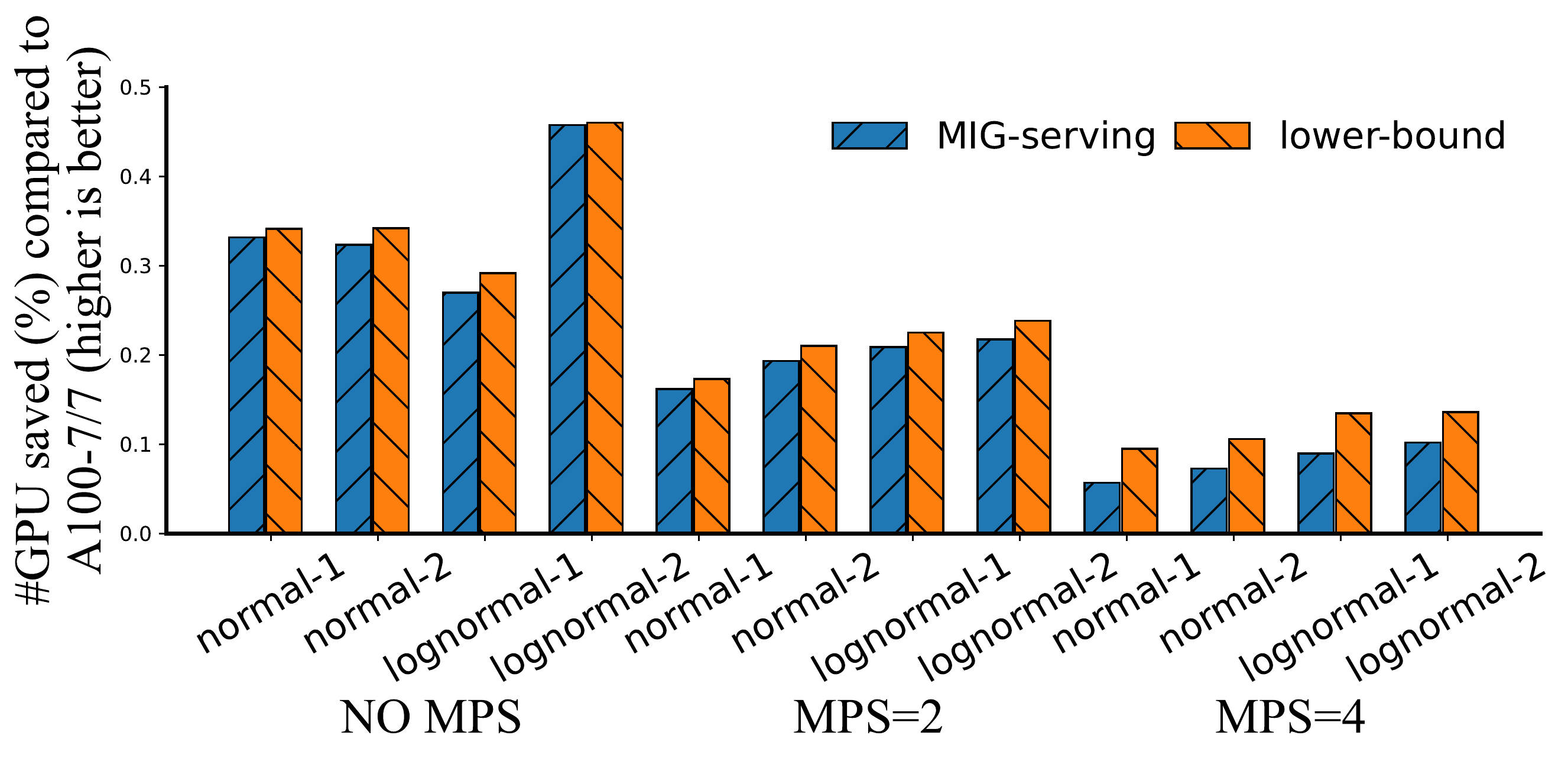}
\end{center}
  \vspace{-3ex}
\caption{The ratio of GPU saved compared to using A100 as-is (``A100-7/7'')
with different MPS configurations.
The left four bar clusters represent the GPUs saved with no MPS;
the middle four represent the configuration with \emph{at most} two MPS processes in each instance;
and the last four with at maximum four MPS processes.
}
\label{fig:mps}
\end{figure}
 
\heading{Combining MIG and MPS.}
MIG and Multi-process service (MPS) are orthogonal techniques that can be used
together. With MPS, multiple processes can share the same GPU, which applies to
GPU instances generated by MIG.
We combine MIG and MPS by running up to $N$ processes of the same model in one GPU instance (e.g., \oneslice).
In our experiments, we use $N=2,\, 4$ because we experienced out-of-memory exceptions when $N>4$.

Figure~\ref{fig:mps} shows the ratio of GPU saved compared to the baseline \gpuoneseven.
By using MPS, multiple processes share GPU resources, and the GPU utilization increases.
Hence, the baseline, \gpuoneseven, has better performance and the GPUs saved by \sys are not as many.
When using four MPS processes, \sys saves about 10\% GPUs.
Nevertheless, MPS increases GPU utilization at the cost of tail latency
stability and failure isolation.
Since MPS has no memory and failure isolation, using it may cause
unexpected high tail latency and correlated failures among model processes.
Deciding whether to use MPS and how many processes to run
is a trade-off that users need to make.

\begin{figure}[t]
\begin{center}
\includegraphics[width=0.4\textwidth]{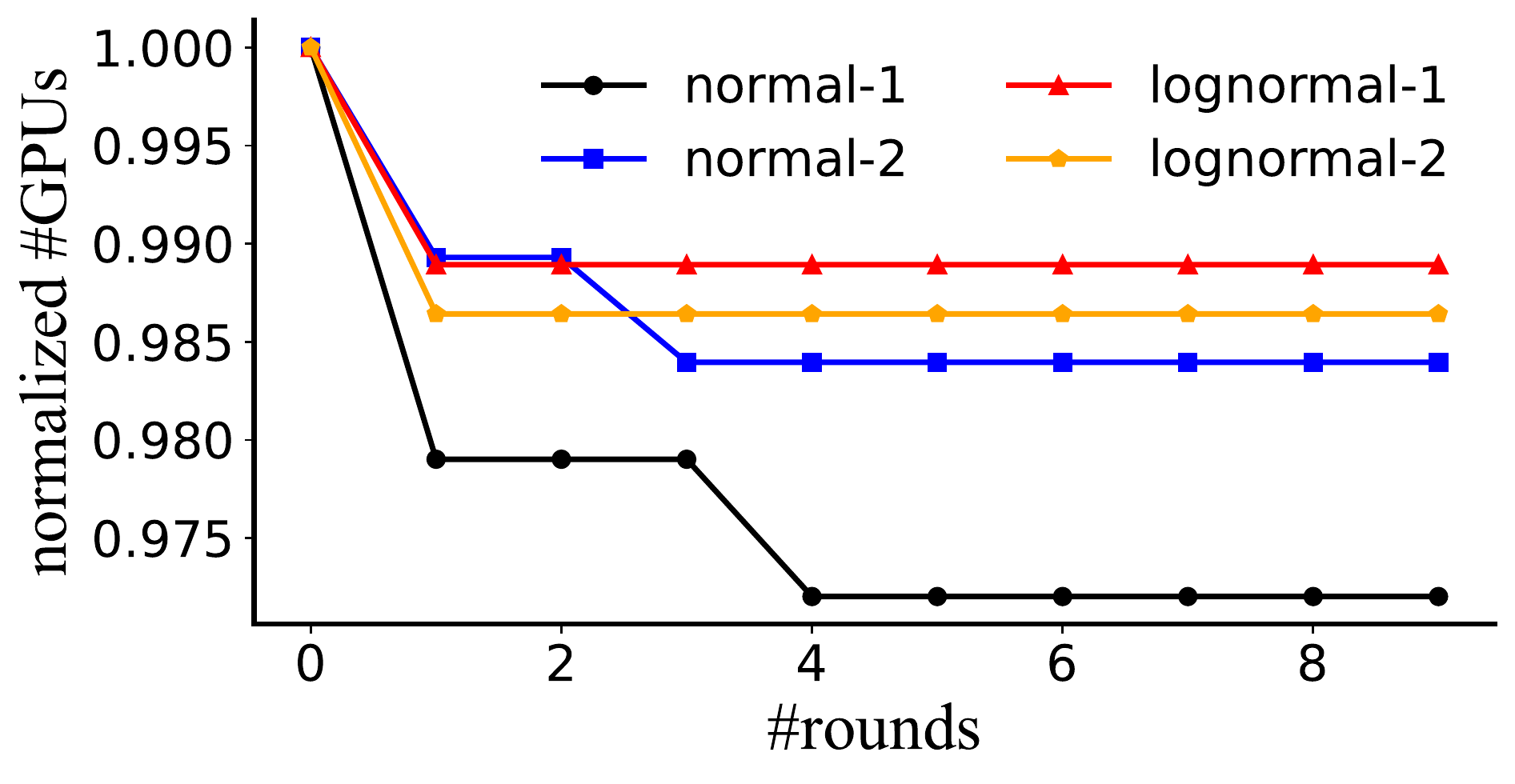}
\vspace{-3ex}
\end{center}
	\caption{The improvement of the slow algorithm (MCTS) over the fast algorithm (heuristic
	greedy) on each GA round. Each line represents a simulation workload.
 The number of GPUs required is normalized to the original \deployment (round 0 of the fast algorithm).
 Notice that the y-axis starts at 0.97.}
\label{fig:twohpase}
\vspace{-3ex}
\end{figure}

\heading{Slow algorithm improvement.}
\label{s:eval:mcts}
We study how the slow algorithm improves over the fast algorithm
by running 10 rounds of GA and MCTS.
Figure~\ref{fig:twohpase} depicts the improvements for each round.
We can see that MCTS improves the solutions of the heuristic greedy algorithm
by saving 1--3\% GPUs,
which is much minor than we expected.
However, it is still worthwhile: we 
can save several to dozens of GPUs
by spending multiple CPU hours.
One of our near future work is to tune GA and MCTS to further improve the \deployments.

\begin{figure*}[htb]
  \begin{subfigure}{0.32\textwidth}
  \includegraphics[width=\linewidth]{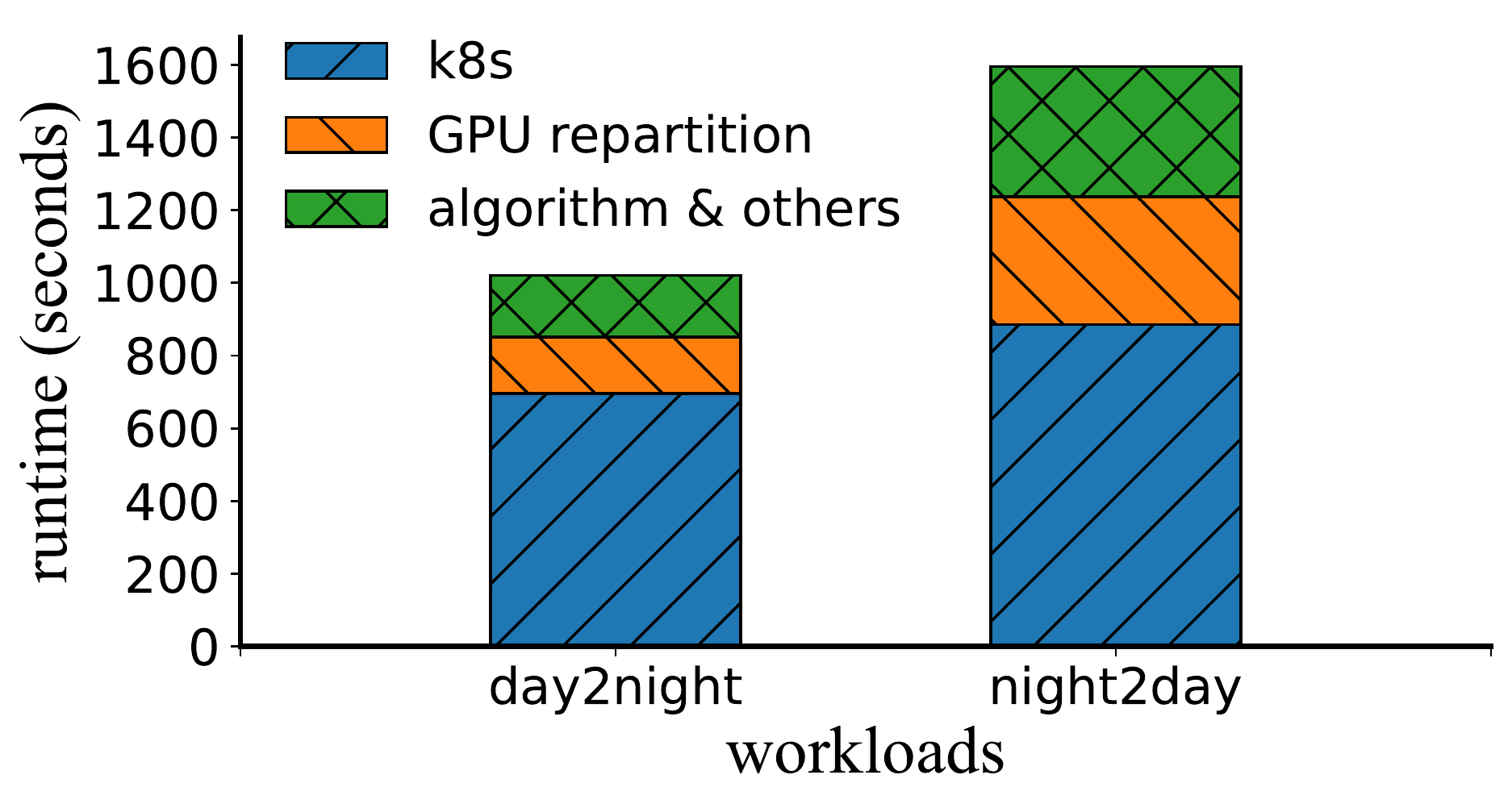}
    \caption{End-to-end runtime of \deployment transitions and time spent on different components.}
  \label{fig:e2estack}
  \end{subfigure}
  \hfill
  \begin{subfigure}{0.32\textwidth}
    \includegraphics[width=\linewidth]{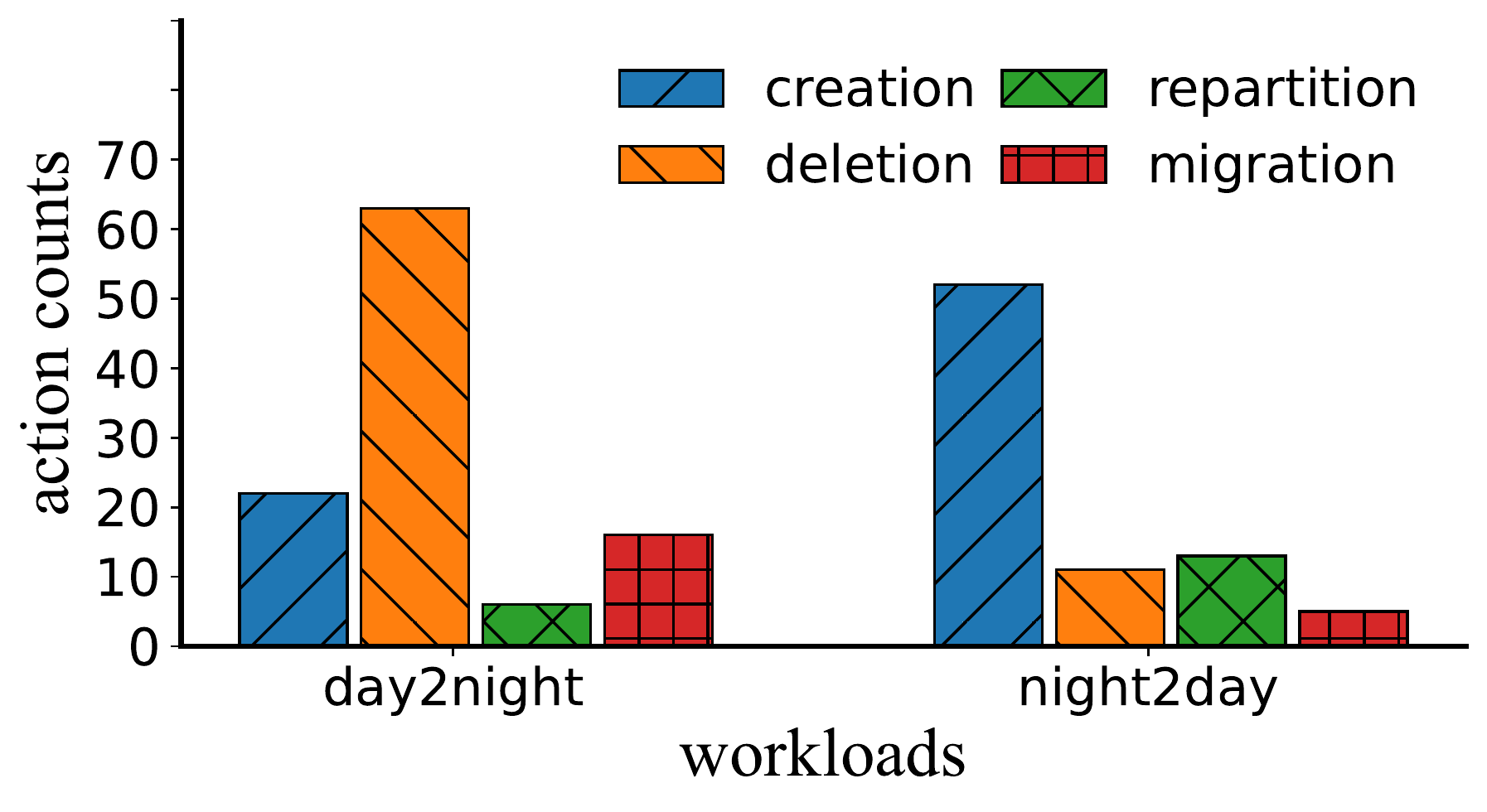}
    \caption{Number of actions during transitions.}
    \label{fig:opcounts}
  \end{subfigure}
  \hfill
  \begin{subfigure}{0.32\textwidth}
    \includegraphics[width=0.9\linewidth]{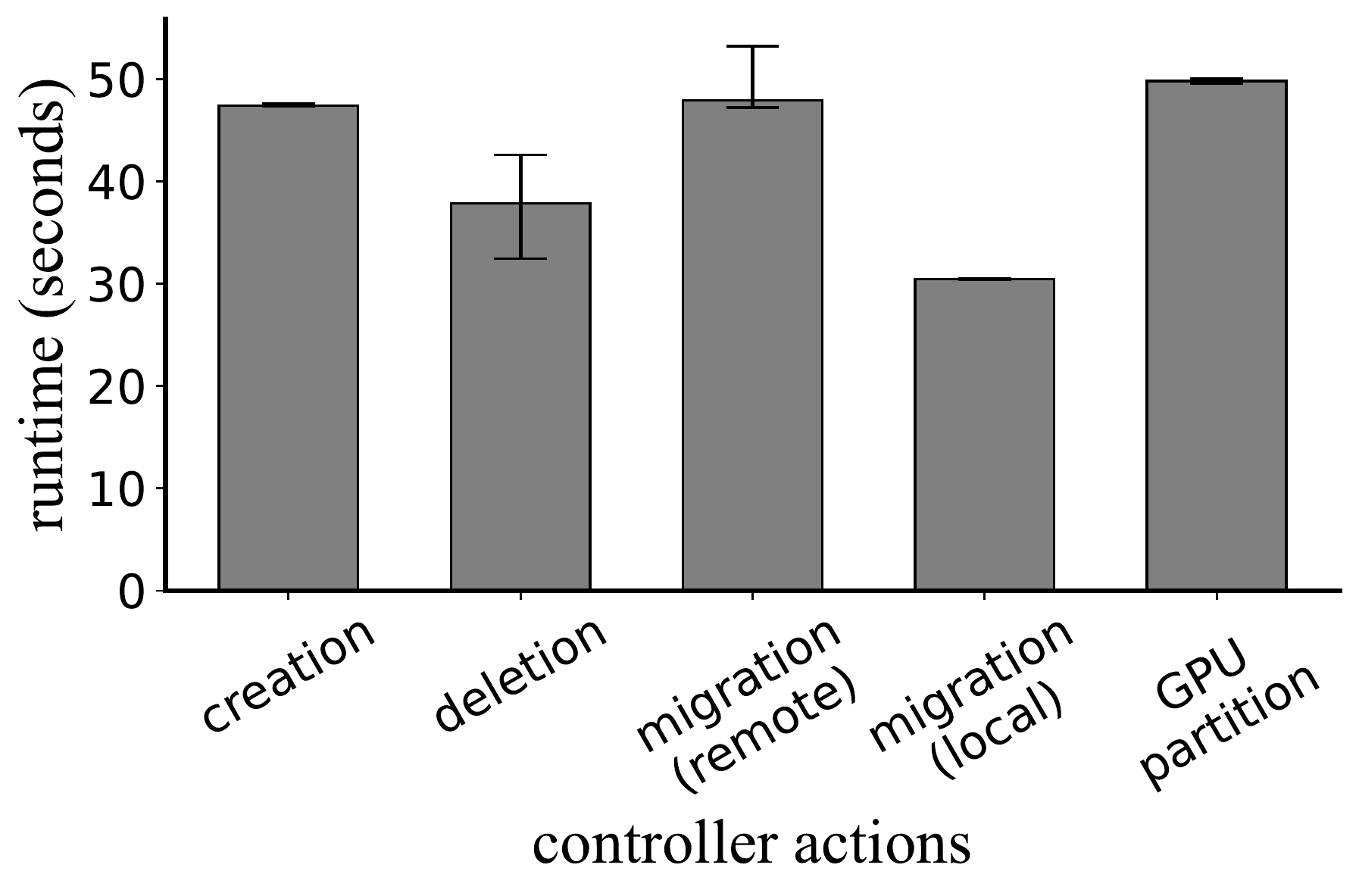}
    \caption{Runtime of (synchronous) \sys actions.
     Bars are average runtime of 10 runs.
     Error bars indicate the maximum and minimum runtime.}
    \label{fig:optime}
  \end{subfigure}
\vspace{-4ex}
\caption{}
\vspace{-3ex}
\label{fig:transition}
\end{figure*}
 
\subsection{\CF{\deployment} transition}
\label{s:eval:transition}

We run \sys on our testbed for the two real-world workloads (the daytime and night workloads),
and experiment with \deployment transitions between the two workloads.
In particular,
we first deploy the five services from the daytime workload, which uses 16 GPUs.
Then, we instruct \sys to switch the \deployment to the night workload, which uses 5 GPUs for the same services.
We call this transition, \textit{day2night}.
Likewise, we call the other way around, \textit{night2day}.

\heading{End-to-end transition runtime.}
We measure the wall-clock time of \sys's two \deployment transitions,
day2night and night2day.
Figure~\ref{fig:e2estack} shows the results.
The transition of day2night is faster than night2day because
the former is mostly shrinking services and reducing the number of GPUs,
whereas the latter requires expanding services and increasing GPUs.

We further decompose each transition runtime into time spent on different components:
k8s, GPU partition, and the \algotwo algorithm.
We find that k8s takes the majority of the time.
By profiling, it turns out that most of k8s' runtime is spent on bootstrapping an
instance (a k8s pod) on certain GPUs.
We believe DNN context switch techniques, like Pipeswitch~\cite{bai20pipeswitch},
can significantly reduce this overhead.

\heading{A closer look at transitions.}
To understand the details of the two transitions,
we record the actions issued by \sys during day2night and night2day,
and summarize them in in Figure~\ref{fig:opcounts}.
The day2night transition issues more instance deletions,
while night2day has more instance creations.
This is because the \deployment during daytime requires more throughputs
(hence instances) 
than the \deployment during night.
Also, night2day has more GPU partition actions because
this transition involves more GPUs,
which need to be configured into the planned partitions.

\heading{\Scheduler actions.}
As mentioned earlier (\S\ref{s:actions}),
\scheduler has four types of actions,
instance creation, deletion, migration (local and remote), and GPU partition.
We measure each of their runtime and show results in Figure~\ref{fig:optime}.
Note that we run these actions in a synchronous manner---we issue an action
and wait until it finishes.
In practice, all these actions are asynchronous and issued in parallel.
\Sys only has to wait when the actions have dependencies,
for example, creating a replacement instance before deleting an old instance.

\subsection{Serving requests in practice}
\label{s:eval:practice}

To understand whether \sys's \deployments satisfy \slos,
we run the two \deployments across models, under the two real-world workloads, on our
testbed and measure their throughputs in practice.
For each \deployment, we run multiple inference clients that continuously issue
requests to DNN services deployed by \sys.
To saturate DNN services, clients gradually increase the number of requests per second
until the throughput reaches its maximum.
In the following, what we report is the maximum throughputs.
Clients and DNN services run on different machines which are connected by 10Gbps datacenter networks.

\begin{figure}[t]
\begin{center}
\includegraphics[width=0.48\textwidth]{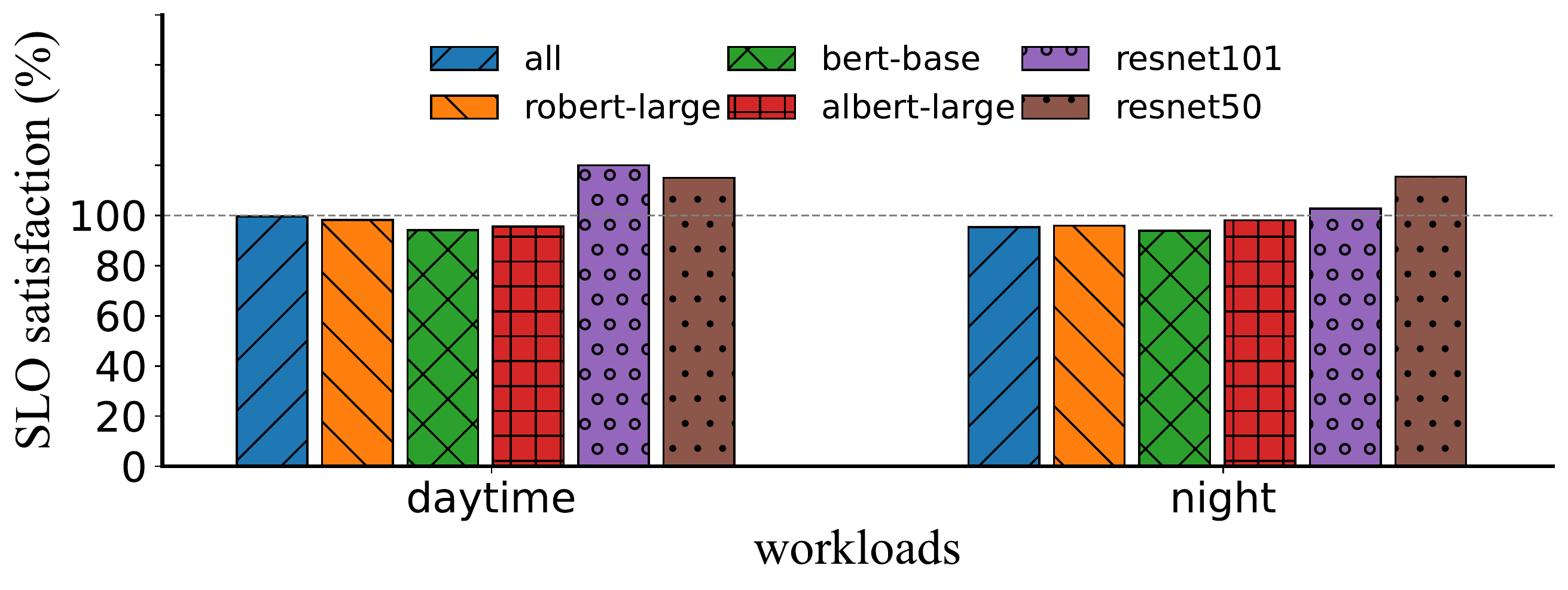}
\end{center}
\caption{Throughputs required by \slos versus throughputs provided by \sys for
the two real-world workloads.
``all'' represents the sum throughputs of all five models.
The y-axis is the ``\slo satisfaction'' (in percentage), which is the throughput provided by \sys
divided by the throughput required by \slos.
The dotted line represents 100\% \slo ssatisfaction.}
\label{fig:practice}
\end{figure}

Figure~\ref{fig:practice} shows the throughputs required by \slos and the
throughputs provided by \sys, for different services.
In general, \sys achieves $>$95\% satisfaction rate for the required throughputs.
The $<$5\% difference is due to the slight performance variance between the model
performance profiling (\S\ref{s:study}, Appendix~\ref{appx:study})
and the performance of serving frameworks (e.g., Tensorflow-serving~\cite{olston17tensorflow}).
This can be improved by collecting model performance in production and
gradually updating profiling data used in \sys's algorithms.

\vspace{-1ex}
\section{Related work}
\label{s:relwork}

\vspace{-1ex}
\heading{Scheduling problems.}
Scheduling problems~\cite{pinedo12scheduling} have been well-studied by multiple communities.
\CF{\prob} is an instance $(R_m |reconf| * )$
that captures DNNs and MIG's characteristics:
(a) model throughputs do not grow linearly with additional resources, %
(b) MIG-enabled GPUs follow specific reconfiguration rules, %
and (c) support partial reconfiguration.

The problem UPM~\cite{pinedo12scheduling} (\textit{Unrelated Parallel Machine Scheduling Problem})
tackles (a),
but it requires a fixed set of machines and does not capture MIG's reconfigurability.
Problems with RMTs~\cite{landers01reconfigurable} (\textit{Reconfigurable Machine Tools})
address (b) and potentially (a).
Examples are FJSSP-CDST~\cite{mahmoodjanloo20flexible} and
permutation flow shop plus RMTs~\cite{azab15modelling}.
But they do not support partial reconfiguration because RMTs have fixed-sized reconfigurable units.
The problem DCSP~\cite{gorczyca09discrete,jozefowska01solving} (\textit{Discrete-Continuous Scheduling Problem})
supports (a) and (c),
but the resources (in our case, GPU slices)
have to be continuous (like power).
But on the contrary, our problem has discrete resources and their allocation is restricted, as indicated by (b).

\vspace{-1ex}
\heading{Partial reconfiguration.}
Similar to MIG-enabled GPUs,
FPGA also supports partial reconfiguration~\cite{fpga_partial}.
A classic reconfigurable device model is the 2D resource
model~\cite{steiger04operating, wang20resource}
which abstracts a job with two constraints, \textit{width} and \textit{height}, representing the
runtime and spatial constraints.
The 2D model targets a similar but different problem:
a job in the model has a fixed width and height,
which in our context means that a service can use one fixed-sized GPU instance.

AmorphOS~\cite{khawaja18sharing} is a system that manages and shares reconfigurable fabric (i.e., FPGA),
and leverages FPGA partial reconfigurations.
AmorphOS has multiple innovations.
The one related to \sys is AmorphOS's scheduler
which uses a two-mode approach (called \textit{low-latency mode} and \textit{high-throughput mode})
to schedule jobs based on the current workloads and FPGA status.
In the context of \prob,
this approach is a rule-based best-effort scheduling algorithm.
Instead of being best-effort,
\sys's algorithms consider wholistically and search in a
large configuration space.

\vspace{-1ex}
\heading{DNN serving systems.}
Traditional DNN serving systems---for example,
Tensorflow-serving~\cite{olston17tensorflow},
TorchServe~\cite{torchserve},
Clipper~\cite{crankshaw17clipper}---mainly focus on optimizing
the performance for a single service instance,
whereas \sys works on a layer below their abstraction:
\sys targets planning GPU instances hence is complementary to them;
these serving frameworks can run within instances created by \sys.

Another thread of DNN serving systems~\cite{gao18low, shen19nexus, gujarati20serving, balasubramanian21accelerating, romero19infaas}
aims at different workloads or specific DNN models.
Though, conceptually, \sys are complementary to these systems
and can run beneath them,
it requires further research to unleash the full potential of both systems.
For example, BatchMaker~\cite{gao18low} improves RNN's inference performance by cellular batching.
Similarly, Nexus~\cite{shen19nexus} accelerates DNN serving %
by batching requests partially across different DNN models.
Because the acceleration of these systems depends on how much portion of requests can be batched,
serving performance varies with workloads.
This is a challenge for \sys as our algorithms require stable performance as inputs.
Likewise, Clockwork~\cite{gujarati20serving} requires the full control over the
execution stack to achieve predicable inference latency,
whose scheduling decisions may conflict with
\sys \scheduler's. %

\vspace{-1ex}
\heading{DNN training scheduler.}
Scheduling DNN training, especially distributed training,
has been intensively studied~\cite{xiao18gandiva,chaudhary20balancing,gu19tiresias,mahajan20themis,peng18optimus,peng19generic, zhao20hived}.
Though \prob covers the problem of DNN training scheduling,
\sys only tackles DNN serving.
As stated in section~\ref{s:actualproblem},
the problem of scheduling DNN inference services
is simpler because DNN services are long-running.
How to apply \sys's techniques to training is our future work.

Though having different contexts, %
some training systems
use related techniques to \sys.
Gavel~\cite{narayanan20heterogeneity} encodes
the problem of training DNNs on heterogeneous GPUs
into an integer programming problem, and uses a solver to solve the problem.
Likewise, \sys's problem can also be expressed in mixed integer programming.
We've tried this,
but our implementation (in Z3~\cite{de08z3})
does not meet our performance requirement---it solves a 5-GPU problem in 20min.
Pollux~\cite{qiao20pollux} uses Genetic Algorithms (GA)
to optimize cluster-wide ``goodput'' (a combination of training throughput and efficiency)
by dynamically re-assigning resources to different jobs.
\Sys also uses GA, but the similarity stops here;
the two systems have different targets
and the contents of GA operations are different.

\vspace{-1ex}
\heading{Heuristic algorithms.}
Many real-world problems are NP-complete,
and in practice people use heuristic algorithms to ``solve'' them.
Our work \Sys shares the same spirit and is indebted to many prior algorithms and systems.
For example, AlphaGo~\cite{silver17mastering} inspires \sys's customized MCTS algorithm (\S\ref{s:slowalgo}).
Similarly, SwapAdvisor~\cite{huang20swapadvisor} enlightens
us to use the Genetic Algorithm in \sys.

\vspace{-1ex}
\section{Summary and future work}

  This paper studies a new hardware feature, MIG, introduced by NVIDIA A100 GPUs.
  To capture the characteristics of MIG and DNN models running on it,
  we introduce a new abstract problem, \prob.
  Further, we design and implement a system, \sys, which
  addresses the problem of serving DNN models with MIG.
  Evaluation shows that \sys can save up to 40\% GPUs versus using A100 as a whole.

MIG is a new hardware feature,
and there are many future works to explore, just name a few:
first, \sys only focuses on serving.
How to apply \sys's techniques to training
is our future work.
Second, \sys's current slow algorithm---MCTS---needs improvement (\S\ref{s:eval:mcts}).
We plan to fine-tune MCTS, or replace MCTS with other heuristic algorithms,
or use (SMT or MIP) solvers to improve the slow algorithm.
Third, \prob (\shortprob) is a new abstract problem, which has the potential
to extend to other reconfigurable devices, such as FPGA.
One of our future work is to comprehensively study what devices
can be abstracted by the \shortprob and how would our algorithms
help in those scenarios.

\begin{flushleft}
\footnotesize
\setlength{\parskip}{0pt}
\setlength{\itemsep}{0pt}
\bibliographystyle{abbrv}
\bibliography{conferences-long-with-abbr,bibs}
\end{flushleft}

\clearpage
\appendix
\section{\Sys algorithms}
\label{appx:algo}

\subsection{Heuristic greedy algorithm}
\label{appx:fastalgo}

As mentioned in section~\ref{s:fastalgo}, \sys uses a greedy algorithm 
as the fast algorithm which is based on a heuristic score.
Figure~\ref{fig:fastalgocode} describes this greedy algorithm in detail.

In section~\ref{s:fastalgo},
we deliberately omitted some technicalities for simplicity.
Specifically, in order to keep the each round of search manageable,
the algorithm only consider mixing two services in one GPU  (Line~\ref{li:mixing2}, Figure~\ref{fig:fastalgocode}).
But, when all services are about to be fully satisfied, the algorithm changes its behavior
by mixing more services in one GPU (Line~\ref{li:mixingmore}, Figure~\ref{fig:fastalgocode}).
This is because two services can no longer saturate a GPU,
and the algorithm needs to pack more services.

\subsection{Customized MCTS}
\label{appx:mcts}

As stated in section~\ref{s:slowalgo},
vanilla MCTS does not work for our problem because of two challenges:
(i) tree node has too many children 
and (ii) the classic MCTS estimation procedure
is slow and inaccurate.
We elaborate how \sys addresses these two challenges below.

For the first challenge,
\sys cuts the space by only having the configurations with top-K scores for a node.
Specifically, for each node (i.e., \sstate),
\sys randomly picks five services which are not fully satisfied
and calculates scores (\S\ref{s:score}) 
for the \gconfigs having these services.
Then \sys chooses the top-K  configurations (K=10 by default) as edges to this node
and generates the corresponding child nodes.

For the second challenge,
\sys develops a fast and accurate estimation
by memoization and randomization.
In particular, \sys's estimation
pre-calculates and caches some good candidates for different types of \sstate.
During the estimation procedure,
\sys maps the current node to a type of \sstate (roughly),
and randomly chooses a child node from the pool of good candidates.
It repeats this step until reaches some leaf node.
Our experiments show that this estimation approach
is about two to three orders of magnitude faster than the classic estimation and is accurate.

\begin{figure}[t]
\begin{center}
\begin{algorithmic}[1]

\newcommand{\varmaxscore}{\t{max\_score}}
\newcommand{\varscore}{\t{score}}
\newcommand{\varbestconf}{\t{best\_conf}}

\Procedure{FastAlgo}{\varcr}
  \State $\t{conf\_set} \gets$ all GPU configs mixing at most 2 services \label{li:mixing2}
  \State $\t{ret} \gets$ empty list

  \While{\vartrue}
    \State $\varmaxscore \gets 0$
    \State $\varbestconf \gets \t{None}$

    \State
    \For {$\varconf \in \t{conf\_set}$} \ \ \ \ // find the best config
      \State $\varscore  \gets$ calculate score for \varconf \ \ \ \ //\S\ref{s:score}
      \If {$\varscore > \varmaxscore$}
        \State \varmaxscore $\gets$ \varscore
        \State \varbestconf $\gets$ \varconf \label{li:chooseconf}
      \EndIf
    \EndFor

    \State
    \State $\t{ret} += \varbestconf$
    \State update \varcr{} according to \varbestconf's \utility
    \If {$\varcr >= [100\%]*len(\varcr)$} \label{li:retcond}
      \State \Return \t{ret}
    \EndIf

    \State
    \For {\serv{i} in all services}
      \If {\serv{i} is almost satisfied} \label{li:mixingmore}
        \State \t{confs} $\gets$ mixing \serv{i} with more services
        \State $\t{conf\_set} += \t{confs}$
      \EndIf
    \EndFor

  \EndWhile
\EndProcedure

\end{algorithmic}
\end{center}
\caption{\Sys's fast algorithm (\S\ref{s:fastalgo}).}
\label{fig:fastalgocode}
\end{figure}

\section{A study of serving performance with MIG}
\label{appx:study}

As mentioned in section~\ref{s:study},
to understand inference performance on different sized instances,
we experiment with 49 open sourced models in 
PyTorch Hub~\cite{pytroch_hub} and TensorFlow Hub~\cite{tensorflow_hub},
and collect model inference throughputs and latencies.
This section describes the experiment details
and provides results from more models than the two in section~\ref{s:study}.

\heading{Experimental setup.}
Our testbed is a machine with
AMD EPYC 7742 CPU with 64 cores,
1.96 TB memory,
running
Debian 9 (stretch) OS,
CUDA 11.0.207,
NVIDIA driver version 450.80.02.

In our experiments,
we do not use the serving frameworks like TensorFlow serving or TorchServe
because we want to evaluate inference throughputs and latencies of GPUs only,
without queueing effects or the overheads from these serving frameworks.
Therefore, we develop our own benchmarking tool which
prepares the inputs in memory
and directly call models' inference functions.
The latencies collected by the tool are the running time of model inference on GPUs,
which do not include overheads from inter-process communication or network costs.

\begin{figure*}[t]
\begin{center}
\includegraphics[width=0.95\textwidth]{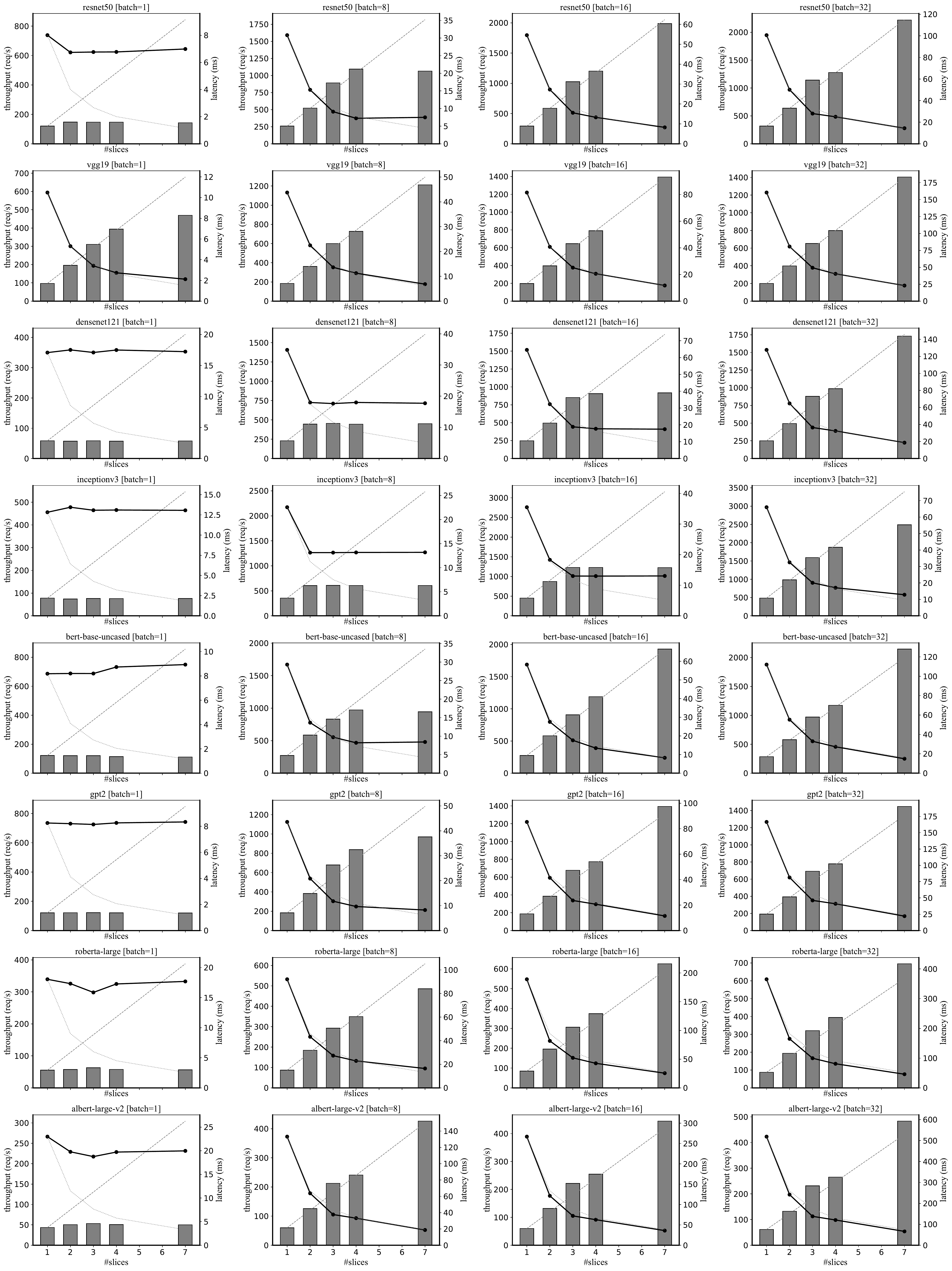}
\end{center}
  \vspace{-3ex}
\caption{PyTorch model inference throughputs and latencies for 1/7--7/7 instances.}
\label{fig:appxpytrochone}
\end{figure*}

\begin{figure*}[t]
\begin{center}
\includegraphics[width=0.95\textwidth]{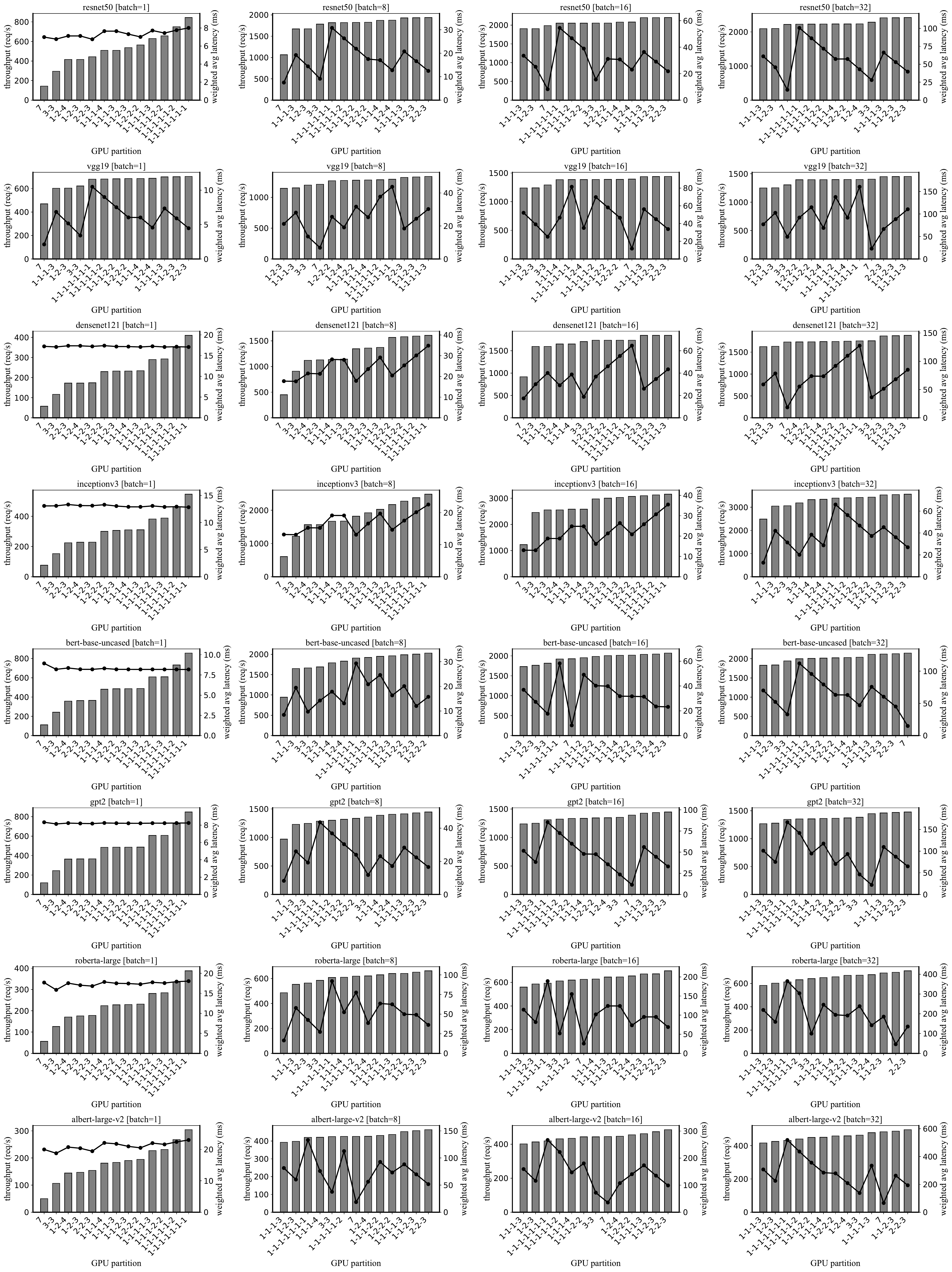}
\end{center}
  \vspace{-3ex}
\caption{PyTorch model inference throughputs and latencies for distinct GPU partitions.}
\label{fig:appxpytrochtwo}
\end{figure*}

\heading{PyTorch.}
We run PyTorch 1.9.0.
Models are fetched from Pytorch Hub.
Figure~\ref{fig:appxpytrochone} shows the single instance throughputs and latencies
for 8 models that exist in both PyTorch's and TensorFlow's model hub (\texttt{resnet50},
\texttt{vgg19},
\texttt{densenet121}, 
\texttt{inceptionv3},
\texttt{bert-base-uncased},
\texttt{gpt2},
\texttt{roberta-large}, and
\texttt{albert-large-v2})
of four batch sizes (1, 8, 16, 32).
Figure~\ref{fig:appxpytrochtwo} shows the throughputs and latencies for different GPU partitions
of the same 8 models.

\begin{figure*}[t]
\begin{center}
\includegraphics[width=0.95\textwidth]{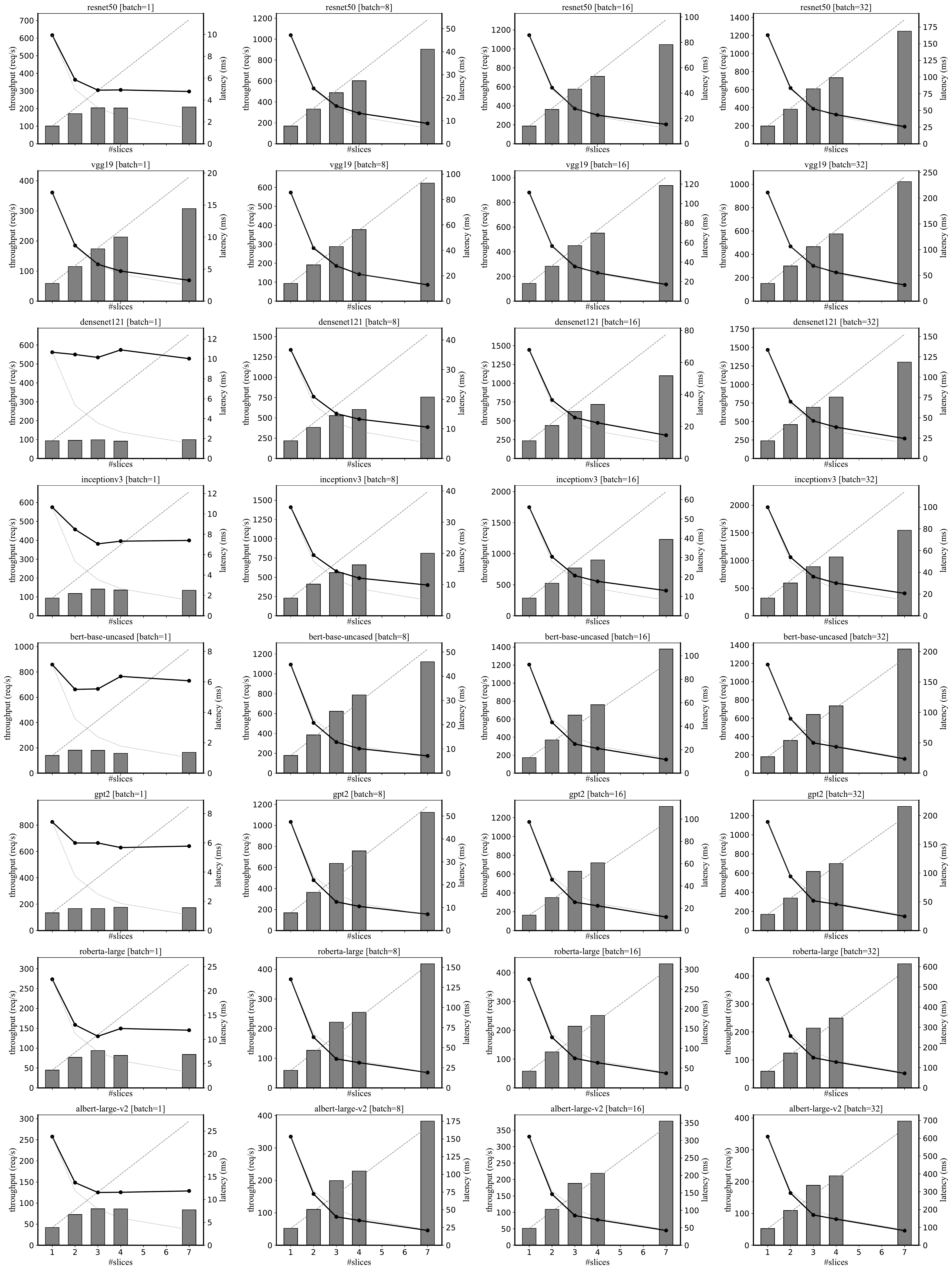}
\end{center}
  \vspace{-3ex}
\caption{TensorFlow model inference throughputs and latencies for 1/7--7/7 instances.}
\label{fig:appxtensorflowone}
\end{figure*}

\begin{figure*}[t]
\begin{center}
\includegraphics[width=0.95\textwidth]{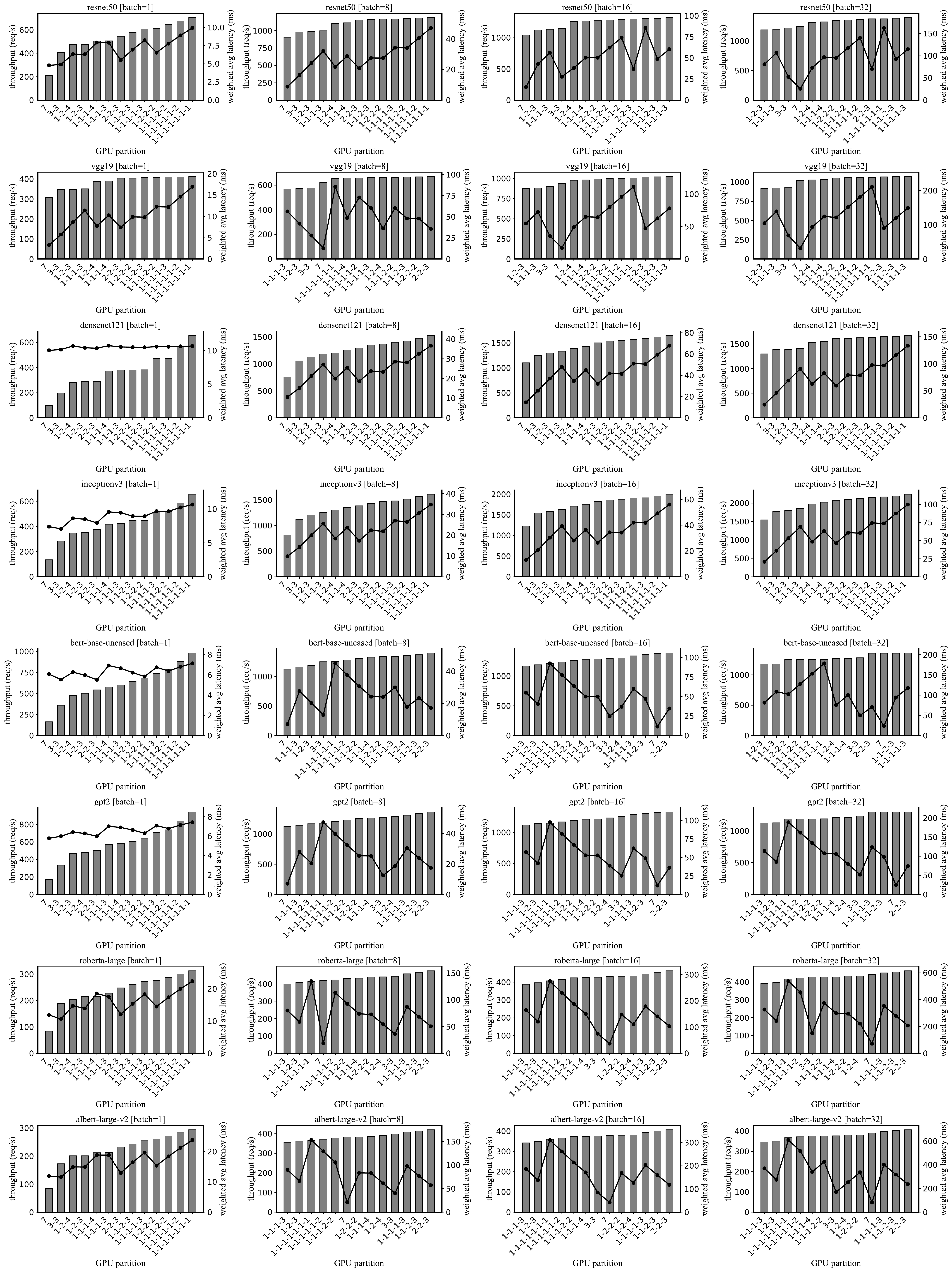}
\end{center}
  \vspace{-3ex}
\caption{TensorFlow model inference throughputs and latencies for distinct GPU partitions.}
\label{fig:appxtensorflowtwo}
\end{figure*}

\heading{TensorFlow.}
We run TensorFlow 2.4.1.
Models are fetched from tensorflow.keras.applications module and transformers 4.5.1 which is a popular Python library for natural language processing.
Figure~\ref{fig:appxtensorflowone} shows the single instance throughputs and latencies
for the same 8 models as PyTorch's above
and in four batch sizes (1, 8, 16, 32).
Figure~\ref{fig:appxtensorflowtwo} shows the throughputs and latencies for different GPU partitions
of the same 8 models.

  
\end{document}